\documentclass{aa}
\usepackage{epsfig} 
\usepackage{graphicx}
\usepackage{amssymb}
\usepackage{natbib}
\usepackage{times}
\def\rxte{{\em RXTE}}
\def\inte{{\it INTEGRAL}}
\def\XMM{{\em XMM-Newton}}

\def\chan{{\em Chandra}}
\def\swift{{\em Swift}}
\def\asca{{\em ASCA}}

\def\igrB{{\em IGR\,J08408-4503}}

\def\xte{{\em XTE\,J1739-302}}

\def\j16479{{\em IGR\,J16479-4514}} 
\def\saxj1818{{\em SAX\,J1818.6-1703}}  

\begin{document}

\title{The supergiant fast X-ray transients XTE\,J1739-302 and IGR\,J08408-4503 
in quiescence with \XMM\ }

\author{E. Bozzo\inst{1}\thanks{email: enrico.bozzo@unige.ch}  
\and L. Stella\inst{2} 
\and C. Ferrigno\inst{1}
\and A. Giunta\inst{2,3} 
\and M. Falanga\inst{4}
\and S. Campana\inst{5}
\and G. Israel\inst{2} 
\and J.C. Leyder\inst{6} 
}

\institute{ISDC - Science Data Centre for Astrophysics, University of Geneva, Chemin d'Ecogia 16, 1290 Versoix, Switzerland. 
\and INAF - Osservatorio Astronomico di Roma, Via Frascati 33, 00044 Rome, Italy. 
\and Dipartimento di Fisica - Universit\`a di Roma Tor Vergata, via della Ricerca Scientifica 1, 00133 Rome, Italy.  
\and International Space Science Institute (ISSI) Hallerstrasse 6, CH-3012 Bern, Switzerland. 
\and INAF - Osservatorio Astronomico di Brera, via Emilio Bianchi 46, I-23807 Merate (LC), Italy. 
\and Institut d'Astrophysique et de G\'eophysique de l'Universit\'e de Li\'ege, 17 All\'ee du 6 ao\'ut, 4000 Li\'ege, Belgium. 
}

\offprints{E. Bozzo}
\titlerunning{XTE\,J1739-302 and IGR\,J08408-4503 in quiescence}
\authorrunning{E. Bozzo et al.}

\abstract{Supergiant fast X-ray transients are a subclass of high mass X-ray binaries  
that host a neutron star accreting mass from the wind of its OB supergiant companion. 
They are characterized by an extremely pronounced and rapid variability in X-rays, which still 
lacks an unambiguous interpretation. A number of deep pointed observations with \XMM\ 
have been carried out to study the quiescent emission of these sources and gain  
insight into the mechanism that causes their X-ray variability.}
{We continued this study by using three \XMM\ observations of the two supergiant 
fast X-ray transient prototypes \xte\ and \igrB\ in quiescence.}
{An in-depth timing and spectral analysis of these data have been carried out.}
{We found that the quiescent emission of these sources is characterized by both complex timing and 
spectral variability, with multiple small flares occurring sporadically after periods of lower 
X-ray emission.   
Some evidence is found in the \XMM\ spectra of a soft component below 
$\sim$2~keV, similar to that observed in the two supergiant fast X-ray transients AX\,J1845.0-0433 
and IGR\,J16207-5129 and in many other high mass X-ray binaries.}
{We suggest some possible interpretations of the timing and spectral properties of the quiescent emission of 
\xte\ and \igrB\ in the context of the different theoretical 
models proposed to interpret the behavior of the supergiant fast X-ray transients.}


\date{Received: 19 January 2010 / Accepted: 09 April 2010}

\maketitle

\section{Introduction}
\label{sec:intro}

Supergiant fast X-ray transients are a subclass of supergiant X-ray binaries 
(SGXBs) that host a neutron star (NS) accreting from the wind of its OB supergiant companion 
\citep[SFXT,][]{sguera05}. In contrast to the previously known supergiant X-ray binaries, SGXBs, 
\citep[i.e., the so-called ``classical SGXBs'', see e.g.,][]{white95}, SFXTs are characterized 
by a pronounced transient-like activity. These sources undergo  
few hours long (as opposed to weeks-months long) outbursts with peak X-ray luminosities 
of $\gtrsim$10$^{36}$~erg/s, and exhibit a 
large dynamic range in X-ray luminosity \citep[$\gtrsim$10$^4$ between outburst 
and quiescence;][]{walter07}. The origin of this extreme variability is  
still debated, and different models have been developed to interpret it. 
One of these models involves a NS accreting matter from   
the extremely clumpy wind of its supergiant companion \citep{zand05, walter07, negueruela08}. 
According to this interpretation, the sporadic capture and accretion of these clumps 
by the compact object can produce the observed fast X-ray flares. 
To reach the required X-ray luminosity swing, very high density 
clumps are required \citep[a factor 10$^4$-10$^5$ denser than the homogeneous stellar wind,][]{walter07}.   
Numerical simulations of supergiant star winds indicate that these high density clumps might 
be produced by instabilities in the wind \citep[][and references therein]{runacres02, oskinova07}.   
\citet{bozzo08} proposed that the X-ray variability of the SFXT sources   
might be driven by centrifugal and/or magnetic ``gating'' mechanisms that can halt most 
of the accretion flow during quiescence, and only occasionally permit direct accretion 
onto the NS \citep[see also][]{grebenev07}. The properties of these gating mechanisms 
depend mainly on the NS magnetic field and spin period. 
At odds with the extremely clumpy wind model, in the gating scenario a transition from the regime in which  
the accretion is (mostly) inhibited to that in which virtually all the captured wind material accretes 
onto the NS requires comparatively small variations in the stellar wind 
velocity and/or density, and can easily give rise to a 
very large dynamic range in X-ray luminosity. 
Yet another model was proposed for IGR\,J11215-5952, so far
the only SFXT to display regular periodic outbursts, which  
envisages that the outbursts take place when the NS in its orbit 
crosses a high density equatorial region in the supergiant's wind \citep{sidoli07}.  

Except for their peculiar X-ray variability, SFXT sources share many 
properties with the previously known SGXBs. 
Measured orbital periods in SFXTs range from 3 to 30~days 
\citep[IGR\,J16479-4514: 3.32~days, IGR\,J17544-2619: 4.92~days; IGR\,J18483-0311: 
18.5~days; SAX\,J1818.6-1703: 30.0~days;][]{jain09,clark09,bird09,zurita09,sguera07}, and 
are thus similar to those of other SGXBs. 
The only exception is the SFXT IGR\,J11215-5952, which has an orbital period of 165~days \citep{sidoli07}. 
The relatively high eccentricities inferred for two SFXTs \citep[$\sim$0.3-0.7;][]{zurita09,rahoui08}
suggest that these systems might be characterized by somewhat more elongated orbits than 
classical SGXBs. 
The spin period of the NS hosted in these sources has been measured only in four cases, the periods
ranging from 4.7 to 228~s \citep[IGR\,J1841.0-0536: 4.7~s; IGR\,J1843-0311: 21 s; IGR\,J16465-4507: 228~s; 
IGR\,J11215-5952: 186.78~s;][]{bamba01, sguera07, lutovinov05, swank07}, and thus being similar to the spin periods measured  
in the classical SGXBs. However, owing to the limited duration of most observations, 
it is not possible to exclude that a number of SFXTs have much longer spin periods 
\citep[see e.g.,][]{smith98,bozzo08}.   

A number of pointed \XMM\ observations of several SFXTs were carried out
to study the quiescent emission of these sources and gain insight into the mechanism that 
drives their peculiar X-ray activity. During the observation of the SFXT IGR\,J16479-4514, \XMM\ captured 
the source undergoing a very steep luminosity decay from the end of an outburst to a much lower state. The latter was 
interpreted in terms of an eclipse of the source by the companion star \citep{bozzo08,bozzo09,jain09}. This observation 
revealed that in at least one case the X-ray variability of a SFXT was due to the obscuration 
by the companion star. 
In the case of IGR\,J18483-0311, \XMM\ helped identify pulsations in the quiescent 
X-ray flux of this source \citep{giunta09}, and thus provided strong support for the idea that SFXTs 
also accrete matter during their quiescent states \citep[see e.g.,][]{sidoli07,bozzo08}. 

To study the low level emission of SFXT sources, 
we present in this paper quiescent state \XMM\ observations of the prototypical SFXTs
XTE\,J1739-302 and IGR\,J08408-4503. 
In Sect.~\ref{sec:sources}, we summarize previous observations of these sources, and in 
Sect.~\ref{sec:data} and \ref{sec:results} we present our data analysis and results. 
In particular, we find that 
\begin{figure*}[t!]
\centering
\includegraphics[scale=0.26,angle=-90]{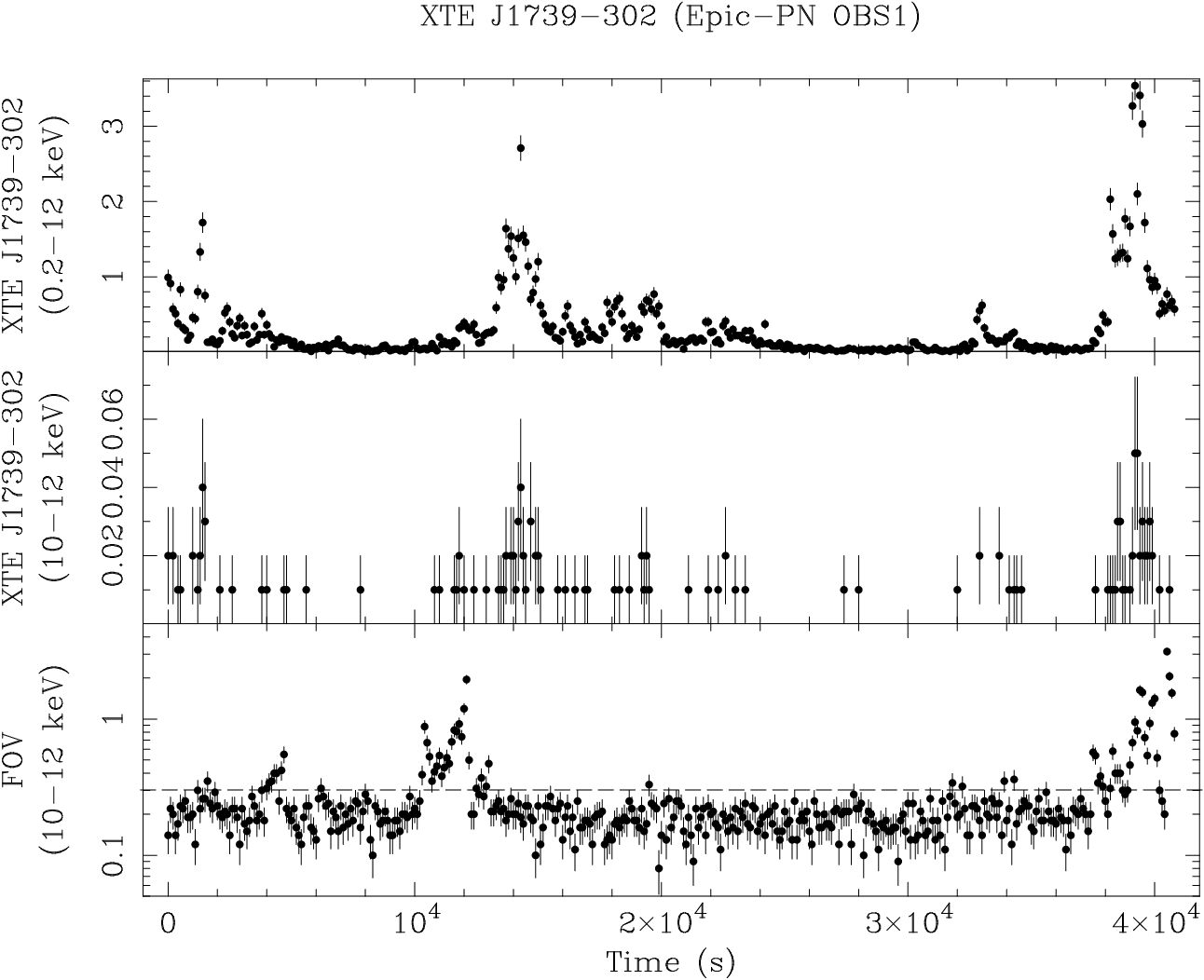}
\includegraphics[scale=0.26,angle=-90]{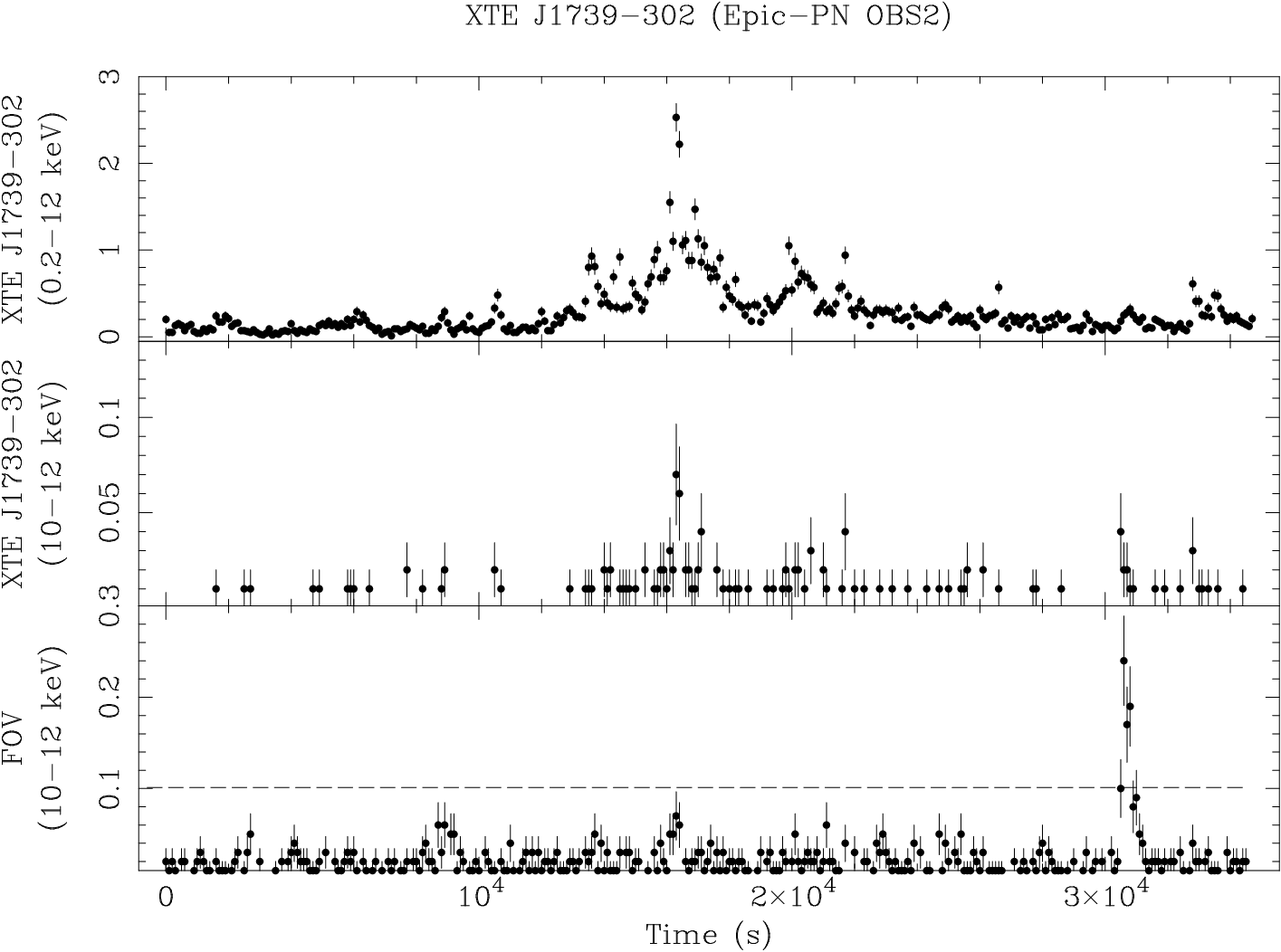}
\includegraphics[scale=0.26,angle=-90]{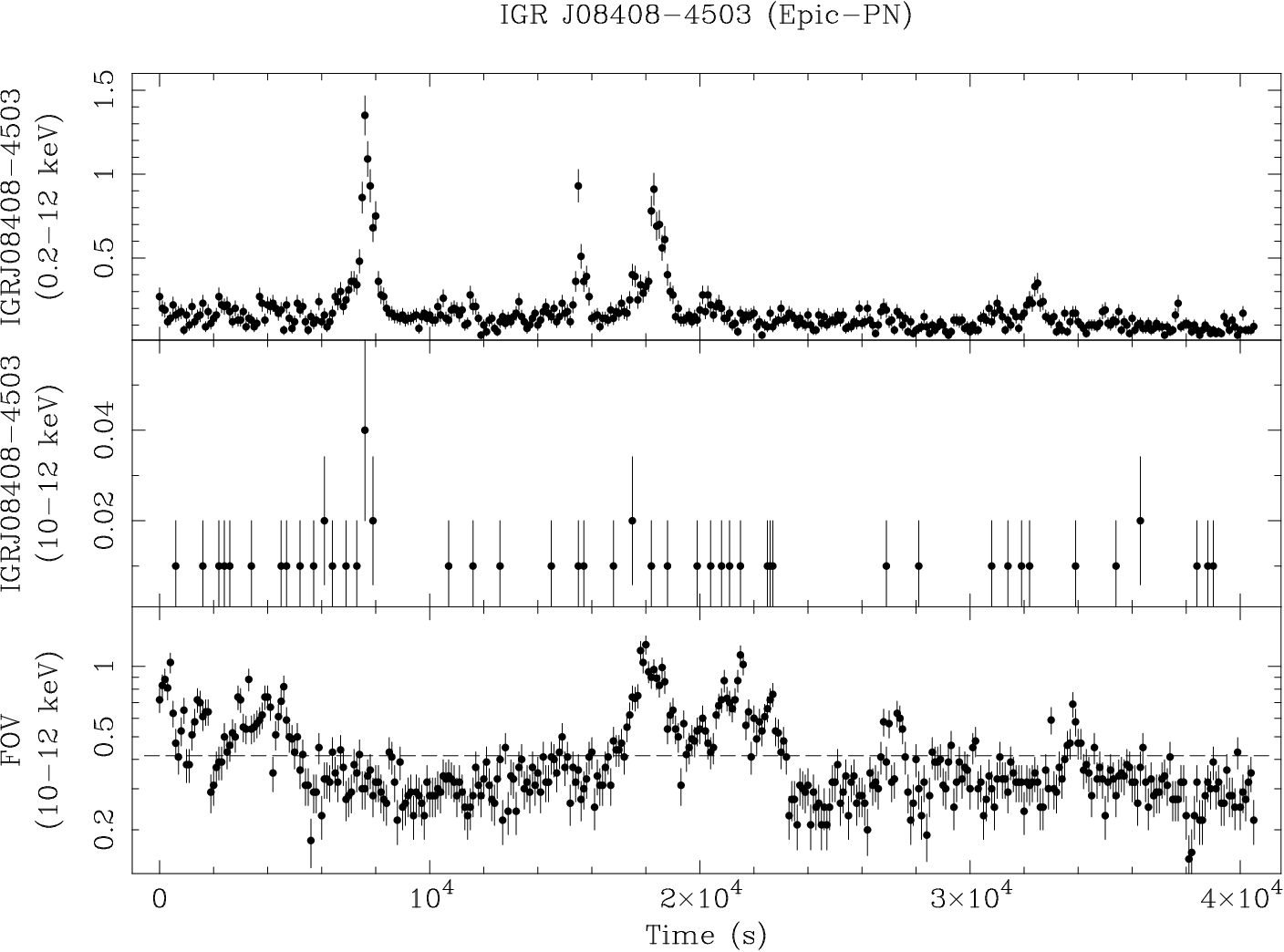}     
\label{fig:backselection} 
\caption{Selection of the high background 
intervals during the two \XMM\ observations (Epic-PN camera) of \xte\ (OBS1 on the 
left and OBS2 in the middle) and the observation of \igrB\ (on the right). 
In each case, we reported the source lightcurve not corrected for the selection of the 
good time intervals and not subtracted for the background in the 0.2-12~keV 
({\it upper panel}) and 10-12~keV energy band ({\it middle panel}), and the 
count rate of the total FOV in the 10-12~keV energy band ({\it lower panel}). 
In all cases, the time bin is 100~s. Only the observational 
intervals in which the total FOV count rate in the 10-12 keV energy band was below 
the threshold indicated with a dashed line were considered for the timing and spectral 
analysis of the sources.}
\end{figure*} 
the quiescent spectra of these sources contain a soft component below 
$\sim$2~keV. 
We discuss some possible
interpretations of this component in Sect.~\ref{sec:discussion}. 
A comparison is also carried out between the quiescent and outburst spectral properties 
of \xte\ and \igrB.\ Our conclusions are summarized in Sect.~\ref{sec:conclusion}.

\section{The sources}
\label{sec:sources}

\subsection{XTE\,J1739-302}
\label{sec:xte} 

\xte\ is a SFXT prototype, and was discovered with \rxte\ during a bright 
outburst in 1997 \citep{smith98}. 
The identification of its supergiant companion led to the determination of the source distance 
at 2.7~kpc \citep{rahoui08}. Several outbursts from this source were detected later with RXTE \citep{smith06}, and 
\inte\ \citep{lutovinov05,sguera05,sguera06,blay08}.  
\xte\ was observed in outburst with \swift\,/BAT on three occasions, on 2008 April 8 \citep{sidoli09}, 
on 2008 August 13 \citep{sidoli09b}, and on 2009 March 10 \citep{romano09}. In only the first 
two cases, \swift\ slewed to the source and observations with the X-ray Telescope, XRT, were carried out. 
During the 2008 April 8 outburst, XRT observed \xte\  
$\sim$387~s after the BAT trigger. These data showed that the source was rapidly ($\sim$1000~s) 
decreasing in intensity, and the X-ray spectrum (0.3-10~keV) could be reproduced well by using an absorbed 
($N_{\rm H}$=13$\times$10$^{22}$~cm$^{-2}$) power law (hereafter, PL) model (photon index $\Gamma$=1.5). The 
0.5-100~keV X-ray luminosity was $\sim$3.0$\times$10$^{36}$~erg/s.  
\citet{sidoli09} also performed an analysis of the \swift\ broad band (0.3-60~keV) 
spectrum of \xte\ during this outburst, 
and found that this spectrum could be reasonably well described by using either a power law with a 
cutoff at high energy ($\sim$13~keV), or a Comptonizing plasma model ({\sc comptt} in {\sc xspec}). 
For the outburst of 2008 August 13, XRT data were obtained starting from $\sim$390~s after the 
BAT trigger, and revealed a more complex behavior than that observed during the previous 
event \citep{sidoli09b}. A time-resolved analysis showed that the source 
X-ray spectrum could be fit equally well by using an absorbed PL 
or a black-body (BB) model with constant photon index or temperature 
($\Gamma$$\sim$1.2, $kT_{\rm BB}$$\sim$1.8~keV), and a varying absorption column density  
(in the range 3-9$\times$10$^{22}$~cm$^{-2}$). The combined XRT+BAT broad band 
(0.3-60~keV) spectrum could be well fit by using either a model of Comptonization of seed photons 
in a hot plasma ({\sc compTT} in {\sc xspec}) or a {\sc BMC} model. 
The BMC comprises a BB component and a component accounting for the Comptonization 
of the BB due to thermal and/or dynamical (bulk) Comptonization. The   
0.1-100~keV X-ray luminosity derived from the simultaneous XRT+BAT spectrum was $\sim$3.8$\times$10$^{36}$~erg/s. 
On 2009 March 10, \xte\ again triggered BAT \citep{romano09}. On this occasion, \swift\ did not 
perform any quick slew towards the source and XRT data were accumulated only $\sim$1.5~h after the BAT trigger. 
At this time, the source was already much fainter (X-ray luminosity $\sim$7$\times$10$^{34}$~erg/s, 2-10~keV), 
and the XRT spectrum could be reproduced well by using an absorbed power-law model 
($N_{\rm H}$=4$\times$10$^{22}$~cm$^{-2}$, $\Gamma$=1.2). 

Little is known about the quiescent emission of \xte.\ An \asca\ observation in 1999 did not detect 
the source and placed a 3~$\sigma$ upper 
limit on its X-ray luminosity of 8$\times$10$^{32}$~erg/s \citep[exposure time $\sim$13~ks,][]{sakano02}. 
A $\sim$5~ks \chan\ observation in 2001 caught the source in a relatively low luminosity state 
(9.7$\times$10$^{33}$~erg/s) and the X-ray spectrum was fit well by 
using an absorbed power-law model \citep[N$_{\rm H}$$\sim$4.2$\times$10$^{22}$~cm$^{-2}$, $\Gamma$=0.62;][]{smith06}. 
Based on a monitoring program with \swift,\ \citet{romano09b} carried out 
the first study of the long-term variation in the quiescent emission from \xte.\ 
They identified three different quiescent states of the source, with a 2-10~keV  
X-ray luminosity of 10$^{35}$, 2$\times$10$^{34}$, and  
5$\times$10$^{32}$~erg/s, respectively. 
The X-ray spectra of these states could be reasonably well described by using an absorbed power-law model, 
with absorption column densities and photon indices in the range (0.3-3.6)$\times$10$^{22}$~cm$^{-2}$ 
and 0.5-1.6, respectively. Alternatively, these spectra could also be fit by using a BB model with temperatures 
and radii in the range 1.3-1.9~keV and 0.02-0.28~km, respectively. With these results at hand, the authors argued  
that the quiescent emission of \xte\ was most likely produced by a spot covering a relatively small region 
of the NS surface, possibly the NS magnetic polar caps. 

\subsection{IGR\,J08408-4503}

\igrB\ was discovered in the Vela region on 2006 May 15 with \inte\ 
during a short flare lasting less than 1000~s \citep{gotz07}. 
Its optical counterpart was later identified as the supergiant 
star HD\,74194 located at 3~kpc, thus confirming that 
this source belongs to the SFXT class \citep{gotz07,masetti06}. 
\igrB\ was observed in outburst two times with \inte,\ and the 
combined JEM-X and ISGRI spectra were most accurately fit by using a rather flat 
cut-off power-law model 
\citep[$\Gamma$$\simeq$0, $E_{\rm cut}$$\sim$11-15~keV,][]{gotz07,leyder08}. 
The absorption column density was 
$\sim$0.1$\times$10$^{22}$~cm$^{-2}$, compatible with the interstellar 
value in the direction of the source \citep{dickey90}.  
The 0.1-100~keV luminosity of the two outbursts was 
7.6$\times$10$^{35}$ and 3$\times$10$^{36}$~erg/s. 

\igrB\ was also caught in outburst by \swift\ 4 times between 2006 and 2009 
\citep{romano08,sidoli09b,barthelmy09}. During the first outburst, 
which occurred on 2006 October 4, \swift\,/XRT slewed to the source $\sim$2000~s 
after the BAT trigger. The combined XRT+BAT spectrum could be reproduced well by using 
a cut-off power-law model ($\Gamma$=0.31, $E_{\rm c}$$\simeq$11~keV). 
The column density was $\sim$0.3$\times$10$^{22}$~cm$^{-2}$. 
The second outburst, which took place on 2008 July 5, was characterized by complex 
behavior and comprised two different flares separated by $\sim$10$^{5}$~s. 
The time-resolved spectral analysis found that the soft X-ray emission of the source 
could be reasonably well described by using an absorbed power-law model with a constant 
photon index of $\sim$1 and a variable absorption column density ranging from 
$\sim$0.5 to $\sim$11$\times$10$^{22}$~cm$^{-2}$. 
The simultaneous XRT+BAT broad band (0.3-80~keV) spectrum of the source 
could be reproduced well using an absorbed cut-off power-law model, with $E_{\rm c}$$\gtrsim$14~keV, 
$\Gamma$$\sim$1.4, and $N_{\rm H}$=6.7$\times$10$^{22}$~cm$^{-2}$. 
The third outburst occurred on 2008 September 21. On this occasion, \swift\,/XRT 
slewed to the source $\sim$150~s after the BAT trigger, and time-resolved 
spectroscopy did not reveal any variation in the absorption column density. 
However, the total XRT spectrum of the observation (exposure time 1160~s) clearly 
revealed a previously undetected soft component below $\sim$2~keV. 
A reasonably good fit to these data could be obtained by using an absorbed 
($N_{\rm H}$=0.4$\times$10$^{22}$~cm$^{-2}$) power-law ($\Gamma$$\simeq$2.2) component 
and a BB at the softer energies (with a temperature of $kT$$\sim$1.95~keV, and a radius of $\sim$1.2~km). 
This BB component was also required to fit the combined XRT+BAT broad band 
spectrum of the outburst (a single comptonization model, BMC, only provided a 
poor fit to the data). 
\citet{sidoli09b} found that the seed photon temperature 
of the BMC component and the BB temperature could not be linked 
to the same value in the fit. This was interpreted as being caused by the presence of 
two distinct photon populations, a colder one with a temperature of about 0.3~keV and a radius of 
$\sim$10~km, and a hotter one with a temperature of 1.4-1.8~keV and a radius of $\sim$1~km. 
However, the statistics of the data did not allow the authors to distinguish which of these two populations 
was seen directly as a BB and which one ended up being seed photons for the thermal Comptonization. 
The latest outburst from \igrB\ was caught by \swift\,/BAT on 2009 August 28. 
Unfortunately, in this case \swift\,/XRT did not perform any follow-up observation, and thus 
no detailed spectral or timing information was available in the 0.3-10~keV energy band. 

To date, the quiescent emission of \igrB\ has remained largely unexplored; the only 
detection of this source during a period of low X-ray activity was obtained 
by two \swift\,/XRT follow-up observations carried out in 2006 May 22 and 
2007 September 29 \citep{leyder08}. 
The exposure time of each of these observations  
was about $\sim$4.0~ks, and only a total of 40 photons could be collected. 
Assuming a Crab-like spectrum, the source X-ray flux was estimated to be 
$\sim$2$\times$10$^{-13}$~erg/cm$^2$/s (0.5-10~keV), corresponding to a luminosity 
of about 2$\times$10$^{32}$~erg/s \citep{kennea06}.

\section{\XMM\ data analysis}
\label{sec:data}

For the present study, we used two \XMM\ observations of \xte,\ and one \XMM\ observation 
of \igrB.\ During all three observations, the two sources were in quiescence. 

\XMM\ observation data files (ODFs) were processed to produce calibrated
event lists using the standard \XMM\ Science Analysis
System (v. 9.0). We used the {\sc epproc} and {\sc
emproc} tasks for the Epic-PN and the two MOS
cameras, respectively. The event files of the two observations 
were filtered to exclude high background time intervals. 
The effective exposure time for each observation and camera is given 
in Sects.~\ref{sec:xteresults} and \ref{sec:igrresults}.  
Source lightcurves and spectra were extracted in the
0.2-15~keV band for the Epic-PN and 0.2-10~keV for the two Epic-MOS 
cameras\footnote{See http://xmm2.esac.esa.int/docs/documents/CAL-TN-0018.pdf.}. 
We extracted the background lightcurves and spectra from the nearest
source-free region to \xte\ and \igrB.\ Background and source circles were all
chosen to lie within the same CCD.  The difference in extraction areas
between source and background was accounted for by using the SAS {\sc
backscale} task for the spectra and the {\sc lcmath} task from
{\sc Heasoft} for the lightcurves. All of the EPIC spectra were rebinned before fitting
so as to have at least 25 counts per bin and, at the same time, prevent oversampling the
energy resolution by more than a factor of three.
Given the low count rate of the source, the Epic-MOS1 
and Epic-MOS2 cameras did not contribute significantly to the spectral analysis; we 
therefore discuss in this paper only the spectra from the Epic-PN camera. 
Throughout this paper, the errors are given at 90\%~c.l. (unless stated otherwise).

\section{Results} 
\label{sec:results}

\subsection{XTE\,J1739-302} 
\label{sec:xteresults} 
\begin{figure}
\centering
\includegraphics[scale=0.4,angle=-90]{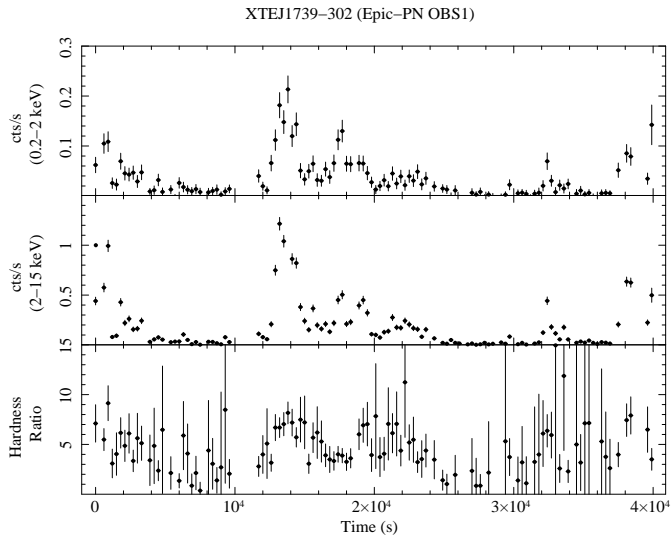}
\caption{ \XMM\ Epic-PN background-subtracted lightcurve of \xte\ during the observation carried out on 
2008 October 1. The upper panel shows the source lightcurve in 
the 0.2-2~keV energy band, whereas the middle panel gives the lightcurve 
in the 2-15~keV energy band (the binning time is 300~s). The ratio of the source count rate in 
the two bands, (2-15~keV)/(0.2-2~keV), versus time is shown in the lower panel. The time intervals in which no data are 
plotted have been discarded because of high background events.}    
\label{fig:xte1} 
\end{figure} 
\XMM\ observed \xte\ twice, on 2008 October 1 (hereafter OBS1) and 
on 2009 March 11 (hereafter OBS2). The Epic-PN camera was operated 
in large window mode in the first case and in small window mode 
during the second observation. 
To identify the high background time intervals, we followed the SAS online analysis 
thread\footnote{See also http://xmm.esac.esa.int/sas/current/documentation/threads/ 
PN\_spectrum\_thread.shtml.} and extracted the Epic-PN lightcurves 
for the full field of view (FOV) in the 10-12~keV energy band. We discarded time intervals in OBS1 
and OBS2 when the 10-12~keV FOV count rate was higher than 0.3 and 0.1~cts/s, respectively. 
This is illustrated in Fig.~\ref{fig:backselection}, where we report the source lightcurves  
(uncorrected for the background and the good time-interval selection) in the energy band 
0.2-12 keV (upper panel), the contribution of the source 
during the two observations in the energy band 10-12 keV (middle panel), and the count rate of the 
total FOV in the energy band 10-12 keV (lower panel). The threshold we imposed in each case is plotted 
with a dashed line in the bottom panel.     
After the good time intervals were selected, we obtained a total effective 
exposure time of 32~ks and 24~ks for OBS1 and OBS2, respectively. 
\begin{figure}
\centering
\includegraphics[scale=0.4,angle=-90]{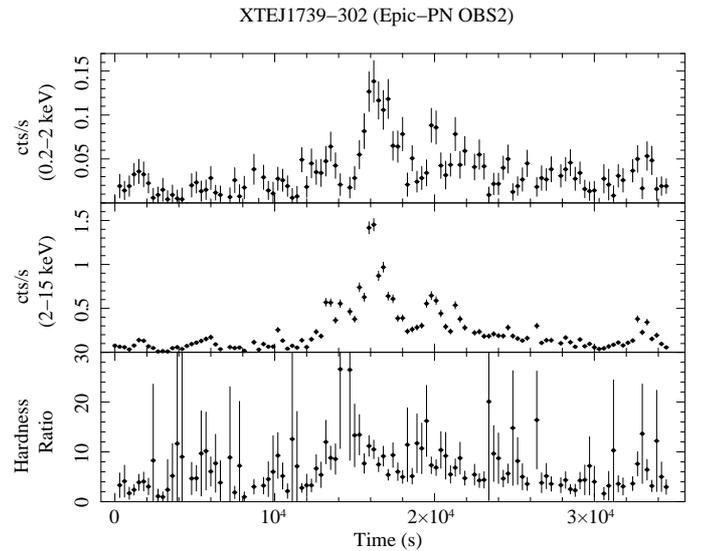}
\caption{The same as Fig.~\ref{fig:xte1}, but for the \XMM\ observation carried out 
on 2009 March 11. The binning time is 300~s.}    
\label{fig:xte2} 
\end{figure}

\begin{figure}
\centering
\includegraphics[scale=0.35,angle=-90]{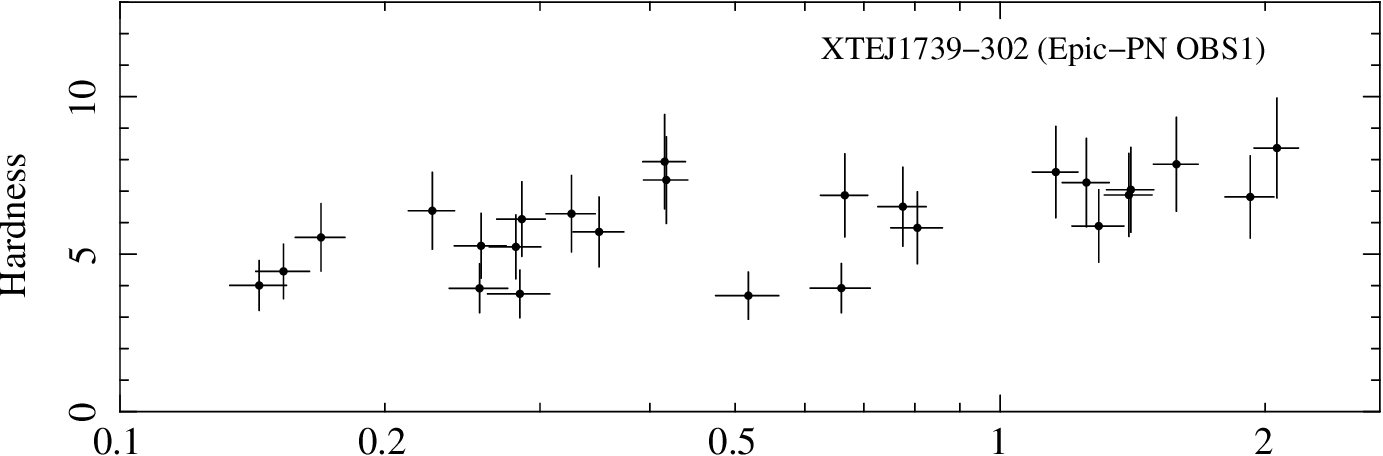}
\includegraphics[scale=0.35,angle=-90]{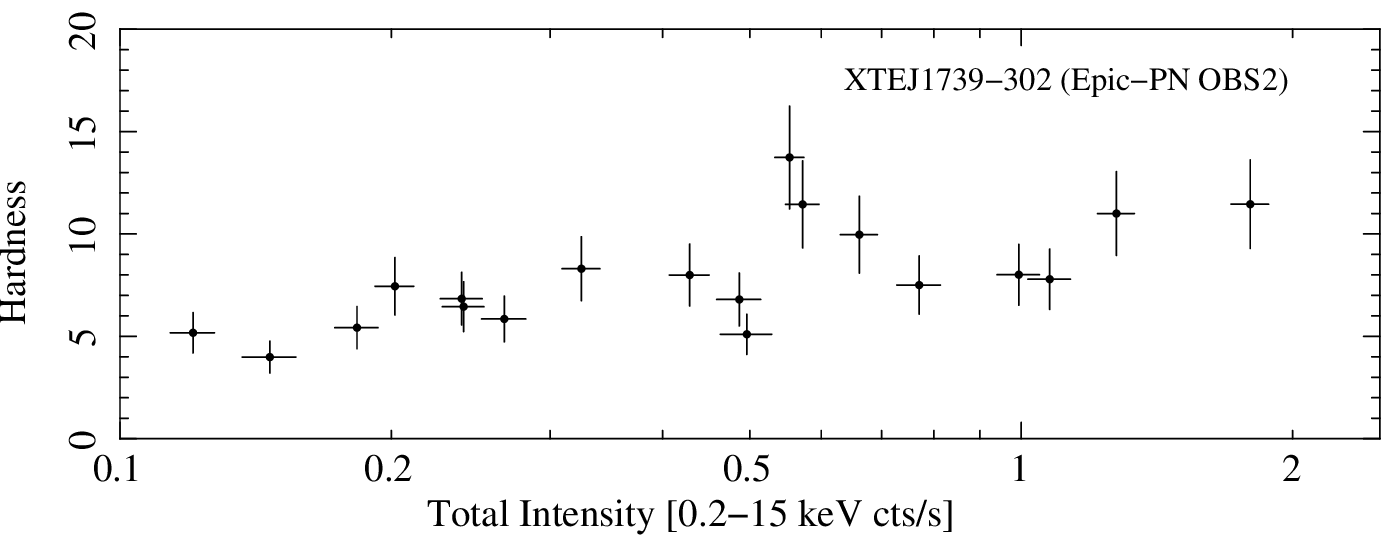}
\caption{Hardness-intensity diagram for the two \XMM\ observations of \xte\ 
(the upper panel is for the observation carried out in 2008, the lower panel for the 
observation in 2009). The hardness is defined as the ratio of the count rate in 
the hard (2-15~keV) to soft (0.2-2~keV) energy band.
The points were obtained by using the 
lightcurves given in Figs.~\ref{fig:xte1} and \ref{fig:xte2}, but, wherever necessary, 
consecutive bins were rebinned so as to achieve a S/N$>$5.5 in the hardness ratio.}      
\label{fig:hardnessxte} 
\end{figure}
\begin{figure}
\centering
\includegraphics[scale=0.35,angle=-90]{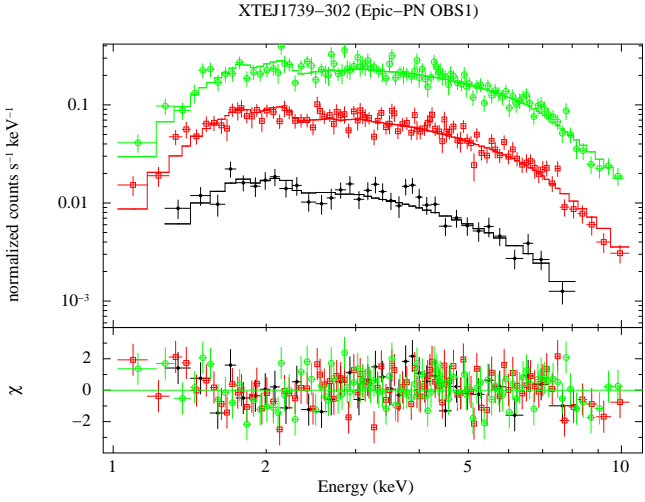}
\caption{The \XMM\ spectra of the \xte\ observation carried out in 2008. 
The open circles, open squares, and filled circles are from the energy resolved data
obtained during the time intervals in which the Epic-PN 0.2-15~keV source 
count rate was $>$0.4, 0.1-0.4, $<$0.1, respectively. The model used for the best 
fits is an absorbed CUTOFFPL (cutoff energy fixed at 13~keV). The residuals from 
this fit are shown in the bottom panel. 
[See the electronic edition of the paper for a color version of this figure.]}     
\label{fig:xtesp1countrate} 
\end{figure} 
\begin{table}[b!]
\centering
\scriptsize
\caption{Results of the  count-rate-resolved spectral analysis for the two \XMM\ observations 
of \xte.\ The model used to fit the data is an absorbed CUTOFFPL (we fixed the cutoff energy 
at 13~keV, see text for details).} 
\begin{tabular}{@{}lllllll@{}}
\noalign{\smallskip} 
\hline
\noalign{\smallskip} 
 & \multicolumn{3}{c}{OBS1} & \multicolumn{3}{c}{OBS2} \\
\noalign{\smallskip}
\hline
\noalign{\smallskip}
cts/s & $<$0.1 & 0.1-0.4 & $>$0.4 & $<$0.1 & 0.1-0.4 & $>$0.4 \\
\noalign{\smallskip}
\hline
\noalign{\smallskip}
$N_{\rm H}$$^a$ &  2.9$_{-0.6}^{+0.5}$ & 2.7$\pm$0.3 &  2.6$\pm$0.3 & 4.1$_{-0.6}^{+0.7}$ & 3.2$_{-0.2}^{+0.3}$ & 3.5$\pm$0.4 \\
\noalign{\smallskip}
$\Gamma$ &  1.8$\pm$0.3 & 1.4$\pm$0.1 & 1.1$\pm$0.1 & 1.8$\pm$0.3 &  1.2$\pm$0.1 & 1.0$\pm$0.1 \\
\noalign{\smallskip} 
$F_{\rm obs}^{b}$ & 4.7$_{-3.2}^{+0.9}$ &  32.2$_{-7.1}^{+3.4}$ & 123.8$_{-20.7}^{+10.0}$ & 9.3$_{-6.9}^{+1.2}$ & 42.0$_{-7.6}^{+4.5}$ & 153.6$_{-26.5}^{+13.7}$ \\
\noalign{\smallskip}
$\chi^2_{\rm red}$  & 1.08 & 1.11 & 1.00 & 0.74 & 1.04 & 1.00 \\
\noalign{\smallskip}
d.o.f.   & 33 & 93 & 105 & 34 & 132 & 89 \\ 
\noalign{\smallskip}
EXP.$^c$ & 18 & 10 & 3 & 10 & 11 & 2 \\
\noalign{\smallskip}
\hline
\noalign{\smallskip} 
\multicolumn{7}{l}{$a$: in units of 10$^{22}$~cm$^{-2}$. $b$: Observed flux in the 0.5-10~keV } \\
\multicolumn{7}{l}{energy band in units of 10$^{-13}$~erg/cm$^2$/s. $c$: Exposure time in ks.} \\
\end{tabular}
\label{tab:xtetimeresolved}
\end{table} 

The lightcurves of the two observations corrected for the good time interval selection and 
background-subtracted are shown in Figs.~\ref{fig:xte1} and \ref{fig:xte2}.  
In both cases, the source displayed a pronounced 
variability on a timescale of hundreds of seconds, with small flares 
occurring sporadically after periods of lower X-ray emission. 
During these flares, the X-ray flux typically increased by a factor 
of $\sim$10-30 (from a few 10$^{-13}$~erg/cm$^2$/s up 
to $\sim$10$^{-11}$~erg/cm$^2$/s). 
\begin{figure}
\centering
\includegraphics[scale=0.35,angle=-90]{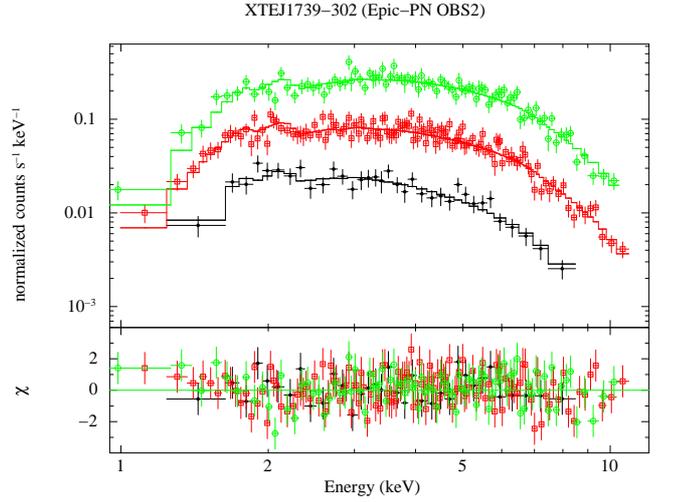}
\caption{The same as Fig.~\ref{fig:xtesp1countrate} but for the observation carried 
out in 2009. The model used for the best fits is an absorbed CUTOFFPL (cutoff energy fixed at 
13~keV). The residuals from these fits are shown in the bottom panel. 
[See the electronic edition of the paper for a color version of this figure.]}     
\label{fig:xtesp2countrate} 
\end{figure}
From the lower panels in Figs.~\ref{fig:xte1} and \ref{fig:xte2}, 
it is apparent that the source emission hardened  
at higher count rates. To investigate this behavior in 
more detail, we plot in Fig.~\ref{fig:hardnessxte} the hardness 
ratio of the source (i.e. the ratio of the source count rate 
in the hard, 2-15~keV, to the soft, 0.2-2~keV, energy bands) as a function 
of the total intensity (i.e. the total source count rate in the 0.2-15 keV energy band). 
For these plots, we rebinned the lightcurves of Figs.~\ref{fig:xte1} 
and \ref{fig:xte2} to obtain a $S/N$$\ge$5.5.   
The scatter in the points in Fig.~\ref{fig:hardnessxte} suggests that 
the relation between the source count rate and the hardness is most likely 
changing from flare to flare. However, an overall increase in the hardness 
with the source intensity is apparent from the plots.   
To test this correlation, we performed a linear regression that accounts
for the errors in both variables \citep{press02}. 
The slope of the linear function determined from this analysis is $1.9\pm0.4$ 
and $4.1\pm0.8$ 
for OBS1 and OBS2, respectively (errors are at 1~$\sigma$ c.l.), 
thus confirming that the correlation is significant ($\sim$5~$\sigma$). 
To investigate which spectral parameters varied with source intensity, we  
divided the lightcurves of the two observations into three intensity levels, and 
accumulated the spectra during the corresponding time intervals. 
We selected the intervals in which the 0.2-15~keV Epic-PN 
count rate of \xte\ was $<0.1$, 0.1-0.4, and $>$0.4, for both OBS1 and OBS2. 
The three spectra extracted from each observation were then fitted together 
with an absorbed cutoff power-law model (hereafter  CUTOFFPL, 
see Figs.~\ref{fig:xtesp1countrate} and \ref{fig:xtesp2countrate}). 
We fixed the cutoff energy at 13~keV \citep{sidoli09}, and found that this model 
provided a closer fit to the data than a simple absorbed PL model. 
We report the results of the fit with the absorbed CUTOFFPL model 
in Table~\ref{tab:xtetimeresolved} (we also attempted to fit the data with an 
absorbed BB model, but this gave a poorer 
result for some of the rate-resolved spectra, 
with $\chi^2$$\gtrsim$1.2). 
\begin{figure}
\centering
\includegraphics[scale=0.35,angle=-90]{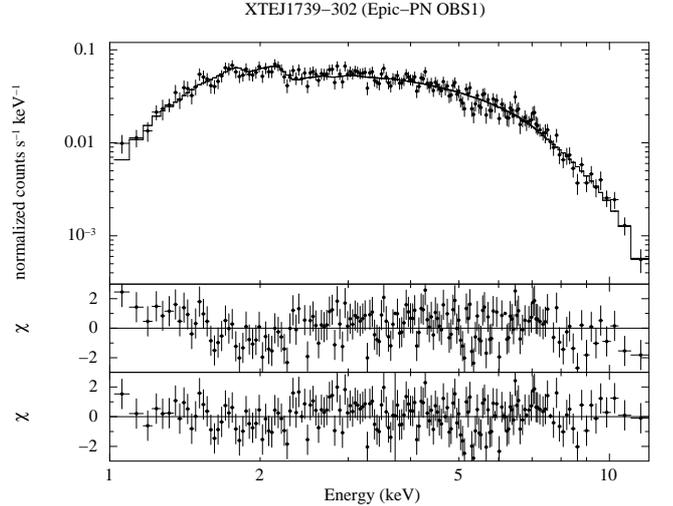}
\caption{The \XMM\ spectrum of \xte\ extracted by using the total exposure time 
of OBS1. The model used for the best 
fit is an absorbed CUTOFFPL (we did not fix here the cutoff energy). The residuals from 
this fit are shown in the bottom panel, whereas the middle panel shows the residuals  
from the fit obtained by using an absorbed CUTOFFPL with the cutoff energy fixed at 13~keV.}     
\label{fig:xtesp1} 
\end{figure}
For OBS1, the rate-resolved spectral analysis
demonstrated that the power-law photon index changed
from 1.1$\pm$0.1, in the highest source count-rate spectrum, to 1.8$\pm$0.3, 
in the spectrum corresponding to the lowest source count rate. 
No significant variation in the absorption column density was measured,    
and the best fit value of $N_{\rm H}$ was consistent (to within the errors) 
with being constant at a level of $\sim$2.6$\times$10$^{22}$~cm$^{-2}$. 
A similar analysis carried out for OBS2 yielded very similar 
results. In this case, the power-law photon index  
decreased at the higher source count rate from 1.8$\pm$0.3 to 
1.0$\pm$0.1, and the absorption column density remained constant 
(to within the errors) at a value of $\sim$4.0$\times$10$^{22}$~cm$^{-2}$. 
\begin{figure}
\centering
\includegraphics[scale=0.35,angle=-90]{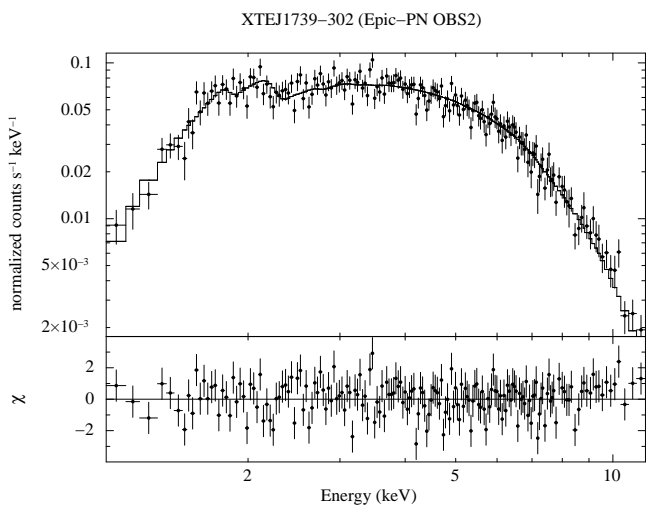}
\caption{The same as Fig.~\ref{fig:xtesp1}, but for the observation carried out 
on 2009 March 11. Here the best-fit model to the data was  
an absorbed CUTOFFPL model (the cutoff energy was allowed to vary during fitting).}     
\label{fig:xtesp2} 
\end{figure}
As the measured changes in the spectral parameters were not dramatic, 
we also extracted the X-ray spectra of the source by using all the data  
from each observation. These spectra are shown   
in Figs.~\ref{fig:xtesp1} and \ref{fig:xtesp2}.
\begin{figure}
\centering
\includegraphics[scale=0.35,angle=-90]{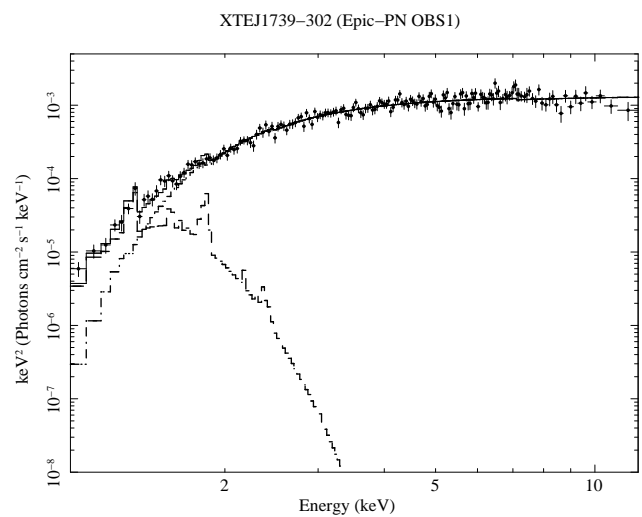}
\caption{An example of an unfolded spectrum of the \XMM\ observation of \xte\ 
carried out in 2008 (OBS1). Here the best fit is obtained by using 
an absorbed CUTOFFPL plus a MKL component (see Sect~\ref{sec:xteresults}).}      
\label{fig:xtesp1unfolded} 
\end{figure} 
A fit to these spectra with a single absorbed PL or BB model 
gave a relatively poor $\chi^2_{red}$ ($\gtrsim$1.3, d.o.f.=175, 184  
for OBS1 and OBS2, respectively) fit, and a slightly better result was obtained  
using the same CUTOFFPL model discussed before for the rate-resolved spectra 
(cutoff energy fixed at 13~keV, $\chi^2_{red}$$\lesssim$1.2, 
d.o.f.=175, 184 for OBS1 and OBS2, respectively). 
However, even in this case, the reduced $\chi^2$ remained significantly 
higher than 1 and, especially in OBS1, some structures are apparent 
in the fit residuals (see Fig.~\ref{fig:xtesp1}). 

By leaving the cutoff energy free to vary in the fit,  
a significantly closer fit was obtained, with a 
$\chi^2_{red}$/d.o.f.=1.02/174, 1.09/183 for OBS1 and OBS2, respectively.
By using the F-test,  
we found that this improvement was highly significant 
for OBS1 (5.3~$\sigma$) and somewhat less significant for OBS2 (3.5~$\sigma$). 
In both cases, the derived cutoff energy turned out to be much lower 
($\sim$4~keV, see Table~\ref{tab:xtefit}) 
than that measured previously when the source was in outburst ($\sim$13~keV). 
Such a low value for the cutoff energy might not be unlikely for 
an X-ray pulsar (see Sect.~\ref{sec:discussion}). 
In the case of OBS1, when an additional spectral component to that of  
the CUTOFFPL appears to be clearly significant, we also tried to investigate the applicability 
of other spectral models. 
\begin{table*}[ht!]
\centering
\scriptsize
\caption{Spectral fits of the \XMM\ data of \xte.\ Below, $N_{\rm H}$ indicates the absorption column 
density, $\Gamma$ the power law photon index, $E_{\rm cut}$ the cutoff energy of the CUTOFFPL component, and 
$R_{\rm BB}$ the BB radius. We indicate with $kT$ the temperature of the BB component, the temperature of the 
COMPTT seed photons, or the temperature of the optically thin gas, depending on the model. 
$kT_{\rm e}$ is the temperature of the Comptonizing electron region in the COMPTT 
model, and $\tau$ its optical depth. We also report the inferred values of 
$N_{\rm MKL}$, which is the normalization of the MKL, $N_{\rm H_2}$, the additional absorption column 
density predicted by the partial covering model, and 0$<$$f$$<$1 the covering fraction. Finally, we give the source flux 
in two different energy bands (0.5-1.5~keV, 1.5-10~keV), and report also the X-ray flux 
in the 0.5-10~keV energy band for completeness.} 
\begin{tabular}{@{}lccccccc@{}}
\hline
\noalign{\smallskip} 
& \multicolumn{5}{c}{OBS1} & & OBS2 \\
\noalign{\smallskip} 
\hline
\noalign{\smallskip} 
Model & {\sc CUTOFFPL} & {\sc BB+CUTOFFPL} & {\sc COMPTT} &  {\sc MKL+CUTOFFPL} & {\sc ZPCFABS*CUTOFFPL} & & CUTOFFPL \\
\noalign{\smallskip}
\hline
\noalign{\smallskip}
$N_{\rm H}$$^a$ & 1.9$^{+0.2}_{-0.3}$ & 1.7$^{+0.4}_{-0.2}$ & 1.3$\pm$0.2 & 3.4$\pm$0.4 & 1.9$\pm$0.4 & & 2.5$^{+0.5}_{-0.2}$ \\
                & (2.7$^{+0.2}_{-0.1}$) & (3.7$\pm$0.4) &  &  &  &  &  (3.4$\pm$0.2) \\
\noalign{\smallskip}
$\Gamma$ & 0.2$^{+0.3}_{-0.4}$ & 0.8$^{+0.4}_{-0.6}$ & --- & 1.5$\pm$0.1  & 1.6$\pm$0.1 &  & 0.1$^{+0.6}_{-0.3}$ \\
                    & (1.3$\pm$0.1) &  (1.5$\pm$0.1) &  &  &  &  & (1.19$\pm$0.07) \\
\noalign{\smallskip}
$E_{\rm cut}$ (keV) & 3.5$^{+0.9}_{-0.7}$ & 13 (fixed) & --- & 13 (fixed) & 13 (fixed) &  & 3.8$^{+2.2}_{-0.6}$                  \\
                    & (13 fixed) &  &  &  &  &  & (13 fixed)  \\
\noalign{\smallskip}
$kT$ (keV) & --- & 1.2$\pm$0.1 & 0.8$\pm$0.1 & 0.15$\pm$0.2 & --- &  & --- \\
                    &  & (0.13$\pm$0.01) &  &  \\ 
\noalign{\smallskip}
$R_{\rm BB}$$^b$ (km) & --- & 0.07$\pm$0.02 & --- & --- & ---  & & ---  \\
                  &  &  (43$_{-22}^{+42}$) &  &  &  \\
\noalign{\smallskip}
$kT_{\rm e}$ (keV) & --- & --- & 2.8$_{-0.7}^{+7.2}$ & --- & --- &  & --- \\
\noalign{\smallskip}
$\tau$ & --- & --- & 6.5$_{-1.6}^{+2.7}$ & --- & --- &  & --- \\
\noalign{\smallskip}
$N_{\rm MKL}$ & --- & --- & --- & 0.3$_{-0.2}^{+1.7}$ & --- & &  --- \\
\noalign{\smallskip}
$N_{\rm H2}$$^c$  & --- & --- & --- & --- & 3.6$_{-1.0}^{+1.5}$ &  & --- \\
\noalign{\smallskip}
$f$      & --- & --- & ---- & ---- & 0.7$\pm$0.1 &  & --- \\
\noalign{\smallskip}
$F_{\rm 0.5-1.5~keV}^{d}$ & 1.7$^{+0.1}_{-0.5}$ & 1.7$^{+0.5}_{-0.7}$ & 1.7$^{+0.6}_{-1.4}$ & 1.7$^{+0.1}_{-0.9}$  & 1.8$^{+0.3}_{-0.5}$ & &  1.1$^{+0.1}_{-0.4}$ \\
                  &  (1.4$\pm$0.2) & (1.7$^{+0.1}_{-1.2}$) &  &  &  &  & (0.8$\pm$0.1)  \\
\noalign{\smallskip}
$F_{\rm 1.5-10~keV}^{e}$ & 2.5$^{+0.2}_{-1.1}$ & 2.5$^{+0.3}_{-1.4}$ & 2.5$^{+0.5}_{-2.3}$ &  2.5$^{+0.3}_{-0.5}$ &  2.5$^{+0.3}_{-0.8}$  &  &  3.9$^{+0.4}_{-1.7}$ \\
                  & (2.5$\pm$0.2) & (2.5$^{+0.2}_{-0.5}$) &  &  &  &  &  (3.9$\pm$0.5)\\
\noalign{\smallskip}
$F_{\rm 0.5-10~keV}^{e}$ & 2.5$^{+0.3}_{-1.0}$ & 2.5$^{+0.3}_{-1.4}$ & 2.5$^{+0.5}_{-2.0}$ &  2.5$^{+0.2}_{-0.5}$ & 2.5$^{+0.3}_{-0.9}$ & &  3.9$^{+0.4}_{-2.0}$ \\
                  & (2.5$\pm$0.3) &  (2.5$^{+0.3}_{-0.5}$)  &  &  &  &  & (3.9$^{+0.4}_{-0.5}$)  \\
\noalign{\smallskip}
$\chi^2_{\rm red}$/d.o.f.  & 1.02/174 & 1.02/173 & 0.98/173 & 1.02/173 & 0.98/173 &  & 1.09/183 \\
                           & (1.20/175) & (0.99/173) &  &  &  &  & (1.16/184) \\
\noalign{\smallskip}
\hline
\noalign{\smallskip} 
\multicolumn{8}{l}{$a$: in units of 10$^{22}$~cm$^{-2}$. $b$: For a distance of 2.7~kpc. $c$: Absorption column density of the 
{\sc zpcfabs} component. $d$: Observed flux in units of 10$^{-14}$~erg/cm$^2$/s.} \\
\multicolumn{8}{l}{$e$: Observed flux in units of 10$^{-12}$~erg/cm$^2$/s.} \\
\end{tabular}
\label{tab:xtefit}
\end{table*}
We again fixed the cutoff energy of the CUTOFFPL component at 13~keV  
and tried first a phenomenological model including an additional black-body component (BB) at 
lower energies ($<$2~keV). We found that both a BB with kT$\sim$1-2~keV and a small emitting area 
(few hundreds m$^2$) and a much colder (kT$\sim$0.1-0.3~keV) BB with an emitting radius of 
$\sim$100~km could reproduce the spectrum reasonably well. 
The interpretation of both of these models for the soft X-ray excess 
faces difficulties: a BB emitting area of about 
one hundred km radius would be much larger than the NS, while a small 
emitting spot on the star surface might also be unlikely in the case of a low-luminosity 
wind-accreting NS (see Sect.~\ref{sec:discussion}).
We therefore tried to fit the spectrum of OBS1 using models 
that provide to alternative physical interpretations. We adopted first a model comprising a    
CUTOFFPL and a MKL component, and then tried a second model in which we included the effect of  
partial covering ({\sc zpcfabs} in {\sc xspec}) on the CUTOFFPL component.   
We note that these two models were suggested to fit the quiescent \XMM\ spectra of the SFXTs AX\,J1845-0433 
\citep{zurita09b} and IGR\,J16207-5129\footnote{This source was classified as an intermediate SFXT by \citet{walter07}.} 
\citep{tomsick09}. In these cases, the authors suggested that the MKL 
component might represent the contribution due to the shocks in the stellar wind 
around the NS, whereas the partial covering may be caused by the obscuration 
of the NS by clumps in this wind. 
Finally, we also attempted to fit the spectrum of OBS1 by using 
the COMPTT and the BMC models, as suggested by \citet{sidoli09} and \citet{sidoli09b}.  
These two models provided a good fit to the data, but in the case of the BMC we found 
that most of the model parameters were only poorly determined in the fit, and thus we do not 
discuss this model further in the case of \xte.\  
The results of the fits obtained with all the other models discussed in this section 
are reported in Table~\ref{tab:xtefit}.  
We checked that a fit to the spectrum of OBS2 with all these models would give results comparable 
to those obtained for OBS1.  
\begin{figure}
\centering
\includegraphics[scale=0.4,angle=-90]{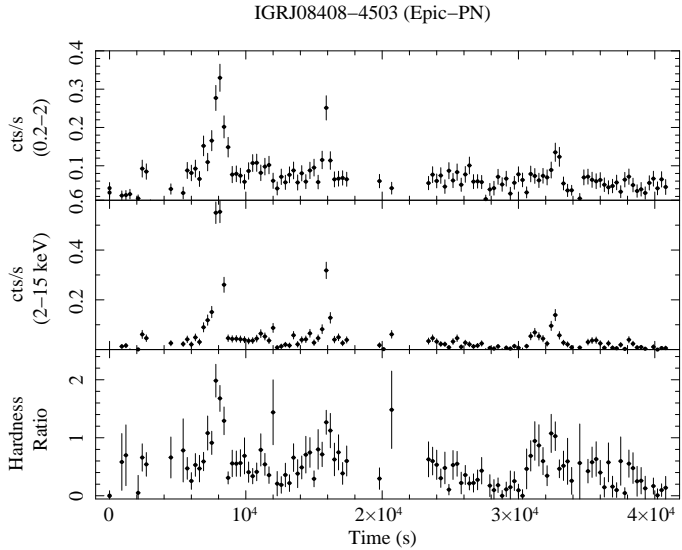}
\caption{ \XMM\ Epic-PN background-subtracted 
lightcurve of \igrB\ during the observation of  
2007 May 29. The upper and middle panel 
shows the source lightcurve in 
the 0.2-2~keV  and 2-15~keV energy bands, respectively.  
The binning time is  300~s. The ratio of the source count rate in 
the two bands versus time is shown in the lower panel.}     
\label{fig:igrb} 
\end{figure} 
All the spectral models reported in Table~\ref{tab:xtefit} provided reasonable values of the fit parameters and 
very similar $\chi^2_{red}$ (note that all the models have the same number of free parameters). 
In the best case, we obtained $\chi^2_{red}$/d.o.f.=0.98/173, and thus the significance of the improvement 
of these fits, with respect to that obtained with a simple absorbed CUTOFFPL model and the energy cutoff 
fixed at 13~keV, was of 5.6~$\sigma$. 
As an example, we show in Fig.~\ref{fig:xtesp1unfolded} the unfolded 
spectrum of OBS1 obtained by using a MKL component to fit the low-energy excess. 
The interpretation of these results is discussed in Sect.~\ref{sec:discussion}. 

We also performed a Fourier analysis of the lightcurves of OBS1 and OBS2, in search 
of coherent pulsations, by using the method described in \citet{israel96}.  
No significant (above 3$\sigma$ level) signal was detected in either observation. 
The corresponding 3$\sigma$ c.l. upper limits
to the pulsed fraction (defined as the semi-amplitude of the sinusoid divided by the source
average count rate), were then computed according to the method described in \citet{vaughan94}.
In OBS1, upper limits at a level of 30\%, 20\%, and 40\% were 
inferred in the 0.1--0.2\,s, 0.2--50\,s, and 50--150\,s period range, respectively. 
In OBS2, we derived upper limits at a level of 20\%, 30\%, and 35\% for periods in the range 
0.03-20~s, 20-50~s, and 0.02-0.03~s, respectively.

\subsection{IGR\,J08408-4503} 
\label{sec:igrresults} 

\XMM\ observed \igrB\ on 2007 May 29, with  
the Epic-PN camera operating in full frame. 
To identify the high background time intervals we 
followed the same technique described for \xte\ 
(see Sect.~\ref{sec:igrresults} and Fig.~\ref{fig:backselection}).  
We extracted the Epic-PN lightcurve  
for the full field of view (FOV) in the 10-12~keV energy band, and set a threshold 
on the full-FOV count rate in this energy band of 0.45~cts/s. 
The total effective exposure time after the good time interval selection for 
\igrB\ was 26~ks. 

From the lightcurve of the observation (Fig.~\ref{fig:igrb}), it is apparent that the variability 
in the quiescent state of this source  
was rather similar to that of \xte.\ In particular, the lower panel of 
Fig.~\ref{fig:igrb} shows that the hardness ratio of \igrB\ increased with the source count rate. 
\begin{figure}
\centering
\includegraphics[scale=0.4,angle=-90]{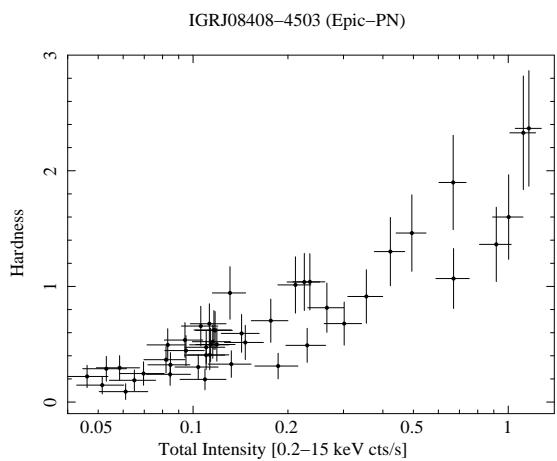}
\caption{Hardness-intensity diagram for \igrB.\ The diagram was created by using 
the 0.2-2~keV and 2-15~keV lightcurves of Fig.~\ref{fig:igrb}, 
but rebinned so as to achieve a S/N$\gtrsim$5.5 in all the hardness ratio values.  
The hardness is defined as the ratio of the count rate in the hard (2-15~keV) 
to soft (0.2-2~keV) energy band.}      
\label{fig:igrbhardness} 
\end{figure}
\begin{figure}
\centering
\includegraphics[scale=0.35,angle=-90]{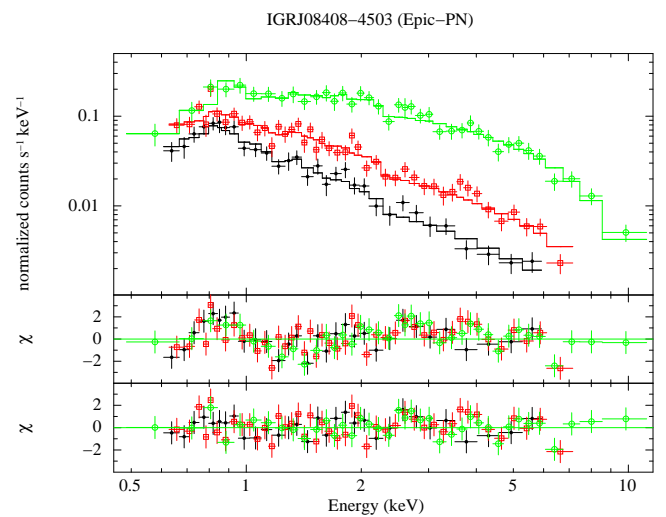}
\caption{The same as Figs.~\ref{fig:xtesp1countrate} and \ref{fig:xtesp2countrate} but for the observation  
of \igrB.\ Here the open circles, open squares, and filled circles refer to  
the spectrum accumulated during the time intervals of the observation in which 
the Epic-PN 0.2-15~keV source rate was $>$0.2, 0.1-0.2, $<$0.1~counts/s respectively. 
The model used for the best fits comprises a MKL plus a CUTOFFPL  
component ($E_{\rm cut}$ fixed at 11~keV). The residuals from these fits are shown 
in the lower panel. The middle panel shows the residuals of the fits obtained by using 
a simple absorbed CUTOFFPL model (cutoff energy fixed at 11~keV). 
[See the electronic edition of the paper for a color version of this figure.]}     
\label{fig:igrbcountrate} 
\end{figure}
\begin{figure}
\centering
\includegraphics[scale=0.35,angle=-90]{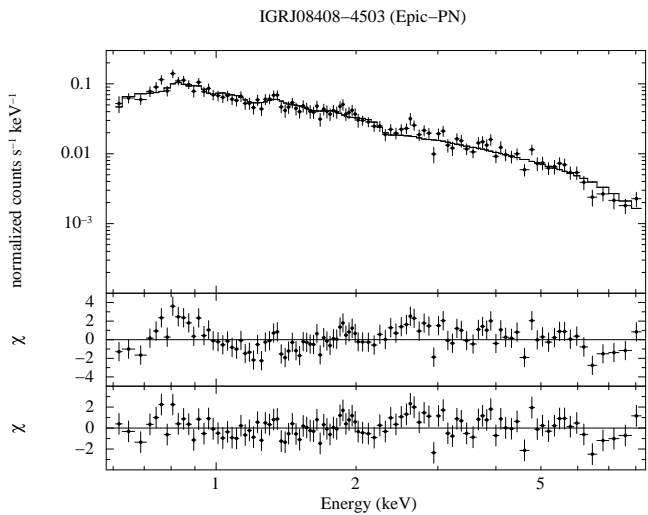}
\caption{The \XMM\ spectrum of \igrB\ extracted by using the total exposure time 
of the observation carried out on 2007 May 29. The model used for the fit is 
an absorbed CUTOFFPL plus a MKL (the cutoff energy is fixed at 11~keV). 
The residuals from the fit are shown in the bottom panel. The middle panel shows the residuals 
of the fit obtained by using only an absorbed CUTOFFPL model (cutoff energy is fixed at 11~keV).}     
\label{fig:igrbsp} 
\end{figure}
\begin{table*}
\centering
\scriptsize
\caption{The same as Table~\ref{tab:xtefit}, but for the time-resolved spectra of \igrB.\ 
In all cases, the cutoff energy is fixed at 11~keV (see Sect~\ref{sec:igrresults}). For the BMC model, we reported 
below the value $\Gamma$=$\alpha$+1, where $\alpha$ is the Comptonization efficiency of the model, the value of the 
``illumination parameter'' $log(A)$, and the normalization $N_{\rm BMC}$.}  
\begin{tabular}{@{}lccccccccc@{}}
\noalign{\smallskip} 
\hline
\noalign{\smallskip} 
Model & \multicolumn{3}{c}{BB+CUTOFFPL} & \multicolumn{3}{c}{MKL+CUTOFFPL} & \multicolumn{3}{c}{BMC} \\
\noalign{\smallskip}
\hline
\noalign{\smallskip}
& $<$0.1 & 0.1-0.2 & $>$0.2 & $<$0.1 & 0.1-0.2 & $>$0.2 & $<$0.1 & 0.1-0.2 & $>$0.2 \\
\noalign{\smallskip}
\hline
\noalign{\smallskip}
$N_{\rm H}$ (10$^{22}$~cm$^{-2}$) & 1.0$^{+0.4}_{-0.3}$ & 0.8$^{+0.2}_{-0.3}$ & 1.2$\pm$0.3 & 0.7$^{+0.1}_{-0.4}$ & 0.6$\pm$0.2 & 0.9$^{+0.2}_{-0.3}$ & 1.1$\pm$0.3 & 0.9$^{+0.2}_{-0.3}$ & 1.4$\pm$0.3 \\
\noalign{\smallskip}
$\Gamma$ & 2.1$^{+0.4}_{-0.3}$ & 1.7$\pm$0.2 & 1.3$\pm$0.2 & 1.7$\pm$0.2 & 1.6$^{+0.1}_{-0.2}$ & 1.1$^{+0.1}_{-0.2}$ & 2.6$^{+0.3}_{-0.2}$ & 2.1$^{+0.2}_{-0.1}$ & 1.7$\pm$0.2 \\
\noalign{\smallskip}
$kT_{\rm BB}$ (keV) & 0.08$\pm$0.02 & 0.08$^{+0.02}_{-0.01}$ & 0.08$^{+0.01}_{-0.02}$ & --- & --- & --- & --- & --- & --- \\
\noalign{\smallskip}
$R_{\rm BB}^{a}$ (km) & 156$_{-46}^{+733}$ & 82$_{-60}^{+160}$ & 286$_{-224}^{+795}$ & --- & --- & --- & --- & --- & --- \\
\noalign{\smallskip}
$kT_{\rm MKL}$ (keV) & --- & --- & --- & 0.22$^{+0.14}_{-0.04}$ & 0.21$^{+0.07}_{-0.04}$ & 0.20$^{+0.10}_{-0.04}$ & --- & --- & --- \\
\noalign{\smallskip}
$N_{\rm MKL}$ (10$^{-3}$)  & --- & --- & --- & 1.0$_{-0.7}^{+2.0}$ & 1.1$_{-1.0}^{+3.6}$ & 7.7$_{-6.8}^{+74.0}$ & --- & --- & --- \\
\noalign{\smallskip}
$kT_{\rm BMC}$ (keV) & --- & --- & --- & --- & --- & --- & 0.07$\pm$0.01 & 0.08$\pm$0.01 & 0.07$\pm$0.01 \\
\noalign{\smallskip}
$log(A)$       & --- & --- & --- & --- & --- & --- & -2.8$^{+0.5}_{-0.2}$ & -2.1$^{+0.4}_{-0.3}$ & -2.9$^{+0.7}_{-0.5}$ \\
\noalign{\smallskip} 
$N_{\rm BMC}$ (10$^{-3}$) & ---  & --- & --- & --- & --- & --- & 4.3$^{+22.3}_{-3.7}$ & 1.1$^{+2.5}_{-1.1}$ & 16.4$^{+75.5}_{-14.4}$ \\
\noalign{\smallskip} 
$F_{\rm 0.5-1.5~keV}^{b}$ & 0.84$_{-0.76}^{+0.05}$ & 1.4$_{-1.0}^{+0.1}$ & 3.0$_{-2.1}^{+0.2}$ & 0.84$_{-0.64}^{+0.12}$ & 1.4$_{-0.9}^{+0.2}$ & 3.0$_{-1.8}^{+0.1}$ & 0.83$_{-0.83}^{+0.12}$ & 1.4$_{-1.4}^{+0.3}$ & 2.9$_{-2.0}^{+0.2}$ \\
\noalign{\smallskip}
$F_{\rm 1.5-10~keV}^{b}$ & 2.2$_{-1.3}^{+0.4}$ & 6.2$_{-2.8}^{+1.1}$ & 38.1$_{-12.0}^{+6.0}$ & 2.3$_{-0.8}^{+0.5}$ & 6.3$_{-2.1}^{+1.0}$ & 38.5$_{-10.1}^{+5.5}$ & 2.2$_{-1.3}^{+0.4}$ & 6.2$_{-6.2}^{+1.4}$ & 38.6$_{-32.6}^{+1.1}$ \\
\noalign{\smallskip}
$F_{\rm 0.5-10~keV}^{b}$ & 3.0$_{-1.9}^{+0.4}$ & 7.6$_{-3.3}^{+1.0}$ & 41.1$_{-17.0}^{+5.0}$ & 3.2$_{-1.5}^{+0.4}$ & 7.7$_{-3.2}^{+1.2}$ & 41.5$_{-12.0}^{+5.0}$ & 2.9$_{-2.7}^{+0.4}$ & 7.7$_{-7.7}^{+1.3}$ & 41.5$_{-29.8}^{+1.1}$ \\
\noalign{\smallskip}
$\chi^2_{\rm red}$/d.o.f.  & 0.8/26 & 1.2/42 & 0.7/38 & 0.8/26 & 1.2/42 & 0.8/38 & 0.8/26 & 1.2/42 & 0.7/38 \\
\noalign{\smallskip}
EXP. (ks)$^{c}$  & 13 & 9 & 2 & 13 & 9 & 2 & 13 & 9 & 2  \\
\noalign{\smallskip}
\hline
\noalign{\smallskip} 
\multicolumn{10}{l}{$a$: For a distance of 3~kpc. $b$: Observed flux in unit of 10$^{-13}$~erg/cm$^2$/s. $c$: Exposure time.} \\ 
\end{tabular}
\label{tab:igrbfit}
\end{table*} 
Figure~\ref{fig:igrbhardness} shows the hardness-intensity 
diagram of \igrB,\ obtained with the same technique 
described in Sect.~\ref{sec:xteresults}. 
In this case, the scatter of the points is somehow less evident than in the case of 
\xte\ and a linear fit to the data required a slope of 1.9$\pm$0.2. 
To investigate the origin of the variability in the hardness ratio of \igrB,\   
we extracted three different spectra during the time intervals of the observation 
in which the source count rate was $>$0.2, 0.1-0.2, and $<$0.1 (hereafter spectra A, B, C). 
A fit to these spectra with a simple absorbed BB or PL model provided unacceptable 
results ($\chi^2_{\rm red}$$\gtrsim$1.5-5.0, d.o.f.=28-44). A CUTOFFPL model with a fixed 
$E_{\rm cut}$=11~keV (see Sect.~\ref{sec:intro}) provided tighter fits to  
the three spectra ($\chi^2_{\rm red}$$\sim$1.2-1.6, d.o.f=28-44).  
However, the $\chi^2_{\rm red}$ was still significantly larger than 1, 
the value of the absorption column density measured from the spectra B and C 
was unreasonably low (compatible with zero), and some structures were 
apparent in the residuals from the fits at energies $<$2~keV (see Fig.~\ref{fig:igrbcountrate}).
A CUTOFFPL model with a free to vary $E_{\rm cut}$ improved only the fit  
to spectrum A ($\chi^2_{\rm red}$/d.o.f=1.03/39), whereas the results of the fits 
to the spectra B and C were almost unchanged.     
To fit the A, B, and C spectra with the same model, and 
obtain at the same time reasonable values of the fit parameters 
(e.g. an $N_{\rm H}$ at least comparable with the Galactic 
value in the direction of the source), 
we assumed that $E_{\rm cut}$=11~keV \citep[as it was found when the source was in outburst,][]{romano08}
and added a second spectral component 
to account for the residuals below $\sim$2~keV. 
We found that a significant improvement in the fits 
could be obtained by adding a MKL or a BB component. 
A single BMC model  
also provided a reasonable fit to the data.   
The results of all these fits are reported in Table~\ref{tab:igrbfit}. 
The significance of the improvement obtained by using these two-component models 
instead of a simple CUTOFFPL is 4, 3, and 4.1~$\sigma$, 
respectively, for the spectra A, B, and C. 

By looking at the results of the fits in Table~\ref{tab:igrbfit}, 
we conclude that the increase in the source hardness ratio with the count 
rate can be most likely ascribed to a variation in the photon index, $\Gamma$,  
rather than a change in the absorption column density or in the properties of the soft component. 
Indeed, the value of $N_{\rm H}$, as well as the temperature and normalization 
of the BB and the MKL components, are compatible with being  
constant (to within the errors) in the three spectra. 
That most of the changes in the 
X-ray flux of the source occurred in the 1.5-10~keV energy band, where the 
contribution of the CUTOFFPL component is much greater than that 
of the soft component (see Fig.~\ref{fig:xtesp1unfolded}), added support 
fot the above conclusion.  
  
\begin{figure}
\centering
\includegraphics[scale=0.35,angle=-90]{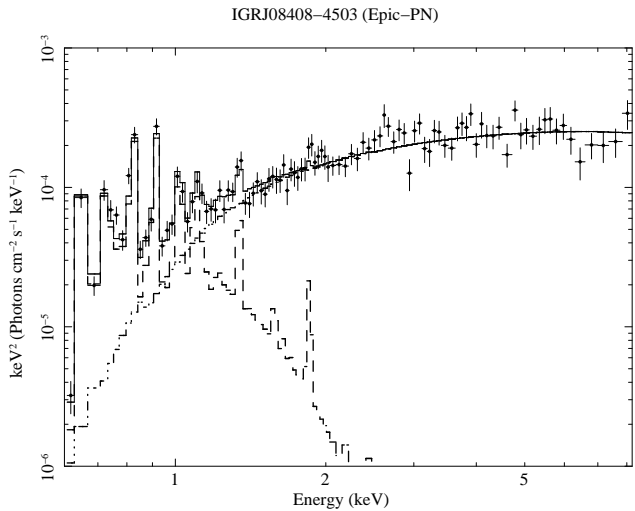}
\caption{The unfolded spectrum of the \XMM\ observation of \igrB.\ 
The best-fit model represented here is obtained 
by using an absorbed CUTOFFPL (cutoff energy fixed at 11~keV) 
plus a MKL component.}      
\label{fig:igrbunfolded} 
\end{figure}

To further constrain the soft component more tightly, we also extracted 
the source spectrum by using the total available exposure time of the \XMM\ observation. 
A fit to this spectrum with a simple BB or a PL model did not provide an 
acceptable result ($\chi^2_{\rm red}$/d.o.f=4.9/101, 1.8/101, respectively 
for the BB and the PL model). A CUTOFFPL model only marginally improved the fit 
($\chi^2_{\rm red}$/d.o.f=1.5/100) and some structures remained present   
in the residuals (see Fig.~\ref{fig:igrbsp}). 
To obtain an acceptable fit to the data, we thus used the same spectral 
models that we adopted for the rate-resolved analysis. All these models provided an equivalently 
good fit to the data. The results of these fits are given in Table~\ref{tab:igrbfittotal} and   
discussed in detail in the following section. We note that when fitting 
the total spectrum of \igrB\ we discarded data in the energy range 0.4-0.6~keV, as we noted that in this energy range 
the background was rather high. Including these points does not affect the best-fit values  
of the model parameters, but indicates that the fit with a CUTOFFPL+BB 
model is slightly preferable ($\chi^2_{\rm red}$/dof=1.0/104) than the  
CUTOFFPL+MKL model ($\chi^2_{\rm red}$/dof=1.1/104). 
Further \XMM\ observations of this 
source are probably needed to resolve this issue.  
In Fig.~\ref{fig:igrbsp}, we show the spectrum of \igrB\ accumulated over 
the entire exposure of the observation and fitted with the MKL+CUTOFFPL model. 
In this figure, we also show for comparison the residuals obtained by fitting the same   
spectrum with a simple absorbed CUTOFFPL model ($E_{\rm cut}$=11~keV, fixed). 
The unfolded spectrum of \igrB\ is shown in Fig.~\ref{fig:igrbunfolded}. 

We searched for pulsations in the power spectra of the \XMM\ observation 
of \igrB\ by using the same technique described in Sect.~\ref{sec:xteresults}. 
No significant (above 3$\sigma$ level) signal was detected in these data. 
We determined an upper limit to the pulsed fraction of 
25\%, 30\%, and 40\% for periods in the range 0.3-50~s, 50-100~s, and 
0.15-0.3~s, respectively (3$\sigma$~c.l.).  

\begin{table}
\centering
\scriptsize
\caption{The same as Table~\ref{tab:igrbfit}, but for the spectrum of \igrB\ obtained by using the total available 
exposure time of the \XMM\ observation.}  
\begin{tabular}{@{}lccc@{}}
\noalign{\smallskip} 
\hline
\noalign{\smallskip} 
Model & BB+CUTOFFPL & MKL+CUTOFFPL & BMC \\
\noalign{\smallskip}
\hline
\noalign{\smallskip}
$N_{\rm H}$ (10$^{22}$~cm$^{-2}$) & 0.9$^{+0.2}_{-0.1}$ & 0.7$^{+0.1}_{-0.2}$ & 1.0$^{+0.1}_{-0.2}$ \\
\noalign{\smallskip}
$\Gamma$ & 1.6$\pm$0.1 & 1.4$\pm$0.1 & 2.0$\pm$0.1 \\
\noalign{\smallskip}
$kT_{\rm BB}$ (keV) & 0.082$^{+0.007}_{-0.004}$ & --- & --- \\
\noalign{\smallskip}
$R_{\rm BB}^{a}$ (km) & 100$^{+126}_{-47}$ & --- & --- \\
\noalign{\smallskip}
$kT_{\rm MKL}$ (keV) & --- & 0.22$^{+0.04}_{-0.02}$ & --- \\
\noalign{\smallskip}
$N_{\rm MKL}$ (10$^{-3}$)  & --- & 1.3$_{-0.8}^{+1.1}$ & --- \\
\noalign{\smallskip}
$kT_{\rm BMC}$ (keV) & --- & --- & 0.075$^{+0.007}_{-0.006}$ \\
\noalign{\smallskip}
$log(A)$       & --- & --- & -2.5$^{+0.3}_{-0.2}$ \\
\noalign{\smallskip} 
$N_{\rm BMC}$ (10$^{-3}$) & ---  & --- & 2.0$^{+2.8}_{-1.3}$ \\
\noalign{\smallskip} 
$F_{\rm 0.5-1.5~keV}^{b}$ & 1.2$_{-0.8}^{+0.1}$ & 1.2$_{-0.7}^{+0.1}$ & 1.2$_{-0.8}^{+0.2}$ \\
\noalign{\smallskip}
$F_{\rm 1.5-10~keV}^{b}$ & 6.3$_{-1.4}^{+0.7}$ & 6.3$_{-1.0}^{+0.7}$ & 6.4$_{-3.7}^{+0.9}$ \\
\noalign{\smallskip}
$F_{\rm 0.5-10~keV}^{b}$ & 7.5$_{-2.1}^{+0.6}$ & 7.5$_{-1.5}^{+0.8}$ & 7.5$_{-4.0}^{+1.1}$  \\
\noalign{\smallskip}
$\chi^2_{\rm red}$/d.o.f.  & 0.99/99 & 1.05/99 & 1.02/99 \\
\noalign{\smallskip}
EXP. (ks)  & 26 & 26 & 26 \\
\noalign{\smallskip}
\hline
\noalign{\smallskip} 
\multicolumn{4}{l}{$a$: For a distance of 3~kpc. $b$: Observed flux in units of 10$^{-13}$~erg/cm$^2$/s.} \\ 
\multicolumn{4}{l}{$c$: Exposure time.} \\ 
\end{tabular}
\label{tab:igrbfittotal}
\end{table}

\section{Discussion} 
\label{sec:discussion} 

We have presented the first deep pointed 
observations of the two prototypical SFXTs, \xte\ and 
\igrB\, in quiescence. 

The two sources exhibited a very complex timing and 
spectral variability, and we discuss them separately 
below. Here we also carry out a comparison between 
their quiescent and outburst emission.   

\subsection{\xte\ }
\label{sec:xtediscussion}

The two \XMM\ observations analyzed in Sect.~\ref{sec:xteresults} 
show that the quiescent emission of \xte\ is characterized by  
a number of low-intensity flares, taking place sporadically from a lower 
persistent emission level. 
The typical duration of these flares is a few thousands of 
seconds, and their X-ray flux is a factor of $\sim$10-30 higher 
than the persistent quiescent flux. 
During the time intervals when the source was at its lowest level of emission, we 
measured an X-ray flux of 4.7$\times$10$^{-13}$~erg/cm$^2$/s (0.5-10~keV). 
This corresponds to a luminosity of 4.1$\times$10$^{32}$~erg/s at 2.7~kpc, and  
is among the lowest values of X-ray luminosity measured from \xte\ 
(a factor of $\sim$2 lower than the 3~$\sigma$ upper limit reported 
by \citet{sakano02}, and comparable with the lowest luminosity 
reported by \citet{romano09b}).  
The total dynamic range of the X-ray luminosity of \xte\ 
between outburst and quiescence is thus $\gtrsim$10$^4$ 
(see Sect.~\ref{sec:intro}).   

The hardness intensity diagrams and rate resolved analysis 
carried out in Sect.~\ref{sec:results} showed that the  
variations in the X-ray flux measured during the \XMM\ observations 
were accompanied by a 
change in the spectral properties of the source,   
the source hardness ratio increasing significantly 
with the count rate. A fit to the rate-resolved spectra with  
a CUTOFFPL model \citep[$E_{\rm cut}$=13~keV][]{sidoli09} implies that 
this behavior originates from a change in the CUTOFFPL 
photon index, $\Gamma$, rather than in a variation 
of the absorption column density.  

These results indicate that the timing and spectral variability of 
\xte\ during the quiescence is qualitatively similar to that observed 
during the outbursts (see Sect.~\ref{sec:intro}). 
A change in the PL photon index with the X-ray flux of \xte\ 
was first noticed by \citet{smith98}. These authors analyzed several different 
flares caught by RXTE, INTEGRAL and ASCA, and showed that the photon index of 
the hard (2-10~keV) X-ray emission from \xte\ changed from  
0.8, when the source X-ray flux was 2.4$\times$10$^{-9}$~erg/s (2-10~keV), to 
2.0, when the X-ray flux was 1.6$\times$10$^{-9}$~erg/s (2-10~keV). 
However, these flares were also characterized by a significant change 
in the absorption column density (from 3 to 37$\times$10$^{22}$~cm$^{-2}$). 
Similar values of the PL photon index and absorption column density were 
reported for the outbursts observed with \swift\ from this source (see Sect.~\ref{sec:intro}).  
In the long-term monitoring of \xte\ carried out with \swift,\ \citet{romano09b} identified 
three different states in the source X-ray flux (Sect.~\ref{sec:intro}), and showed 
that each of these states could be characterized by a different PL photon index 
(in the range 0.5-1.6) and a different absorption column density 
(from 0.3 to 3.6$\times$10$^{22}$~cm$^{-2}$). 
Considering all of these results, it seems unlikely that 
there exists a single monotonic variation in both the absorption column density and 
PL photon index with the source intensity across its quiescent and outburst state. 

When the X-ray spectrum of \xte\ from the entire 
\XMM\ observation is considered, 
the interpretation of the emission from this source becomes even more 
complicated. In particular, 
this spectrum cannot be fit successfully by using a simple absorbed 
PL or a BB model. We indeed showed in Sect.~\ref{sec:xteresults} that these models 
provided only a poor fit to the data, and a more refined model was required. 
Because of the relatively low X-ray flux of the source, different models could 
equivalently describe the data. From a statistical point of view, the  
CUTOFFPL with a variable cutoff energy 
would be preferable, as it provided a very good fit to the data of both OBS1 and OBS2
and requires a lower number of free parameters with respect to the others models 
reported in Table~\ref{tab:xtefit}. 
However, this model would require a cutoff energy much lower ($\sim$4~keV) than the 
value determined when the source was in outburst ($\sim$13~keV). 
This low value for the cutoff energy might not be unlikely for 
an X-ray pulsar (see e.g., the cases of X-Persei and 
RX J0146.9+6121\footnote{Note that these authors used an
absorbed power-law with exponential high-energy cutoff ({\sc highecut*pl} in {\sc xspec}) 
to fit their data.}, \citet{haberl98,disalvo98}, and \citet{coburn02} for a review). 
However, we suggested that, given our relatively poor 
knowledge of the quiescent emission of \xte\ and of SFXTs in general, the applicability of other 
spectral models might be worth exploring.  

Besides the CUTOFFPL, 
the quiescent spectrum of \xte\ might require 
an additional spectral component at the lower energies ($<$2~keV).  
In Sect.~\ref{sec:xteresults}, we showed that a fit to the spectra of  
OBS1 and OBS2 with a CUTOFFPL model and a fixed $E_{\rm cut}$=13~keV
would leave some evident structures in the residuals from the fit and 
the addition of a BB, or a MKL component can significantly improve 
the results. Even if these models require one more free parameter than 
the CUTOFFPL with a variable $E_{\rm cut}$, the results we obtained would then  
be in agreement with those found in the case of \igrB.\ The spectrum 
of this source could, indeed, not be reproduced using a simple 
CUTOFFPL model (see below). 

A soft spectral component below 2~keV might be expected in the spectra 
of the SFXT sources, as this component is very common in binaries 
hosting a NS accreting mass from a massive companion.  
\citet{hickox04} showed 
that the detectability of this component 
depends mainly on the amount of absorption in the direction of the sources. 
According to their results, in the most luminous objects  
($L_{\rm X}$$\simeq$10$^{38}$~erg/s) the soft component is produced 
by reprocessing hard X-rays from the NS by some optically 
thick material (e.g., an accretion disk), whereas for sources with lower 
luminosities ($<$10$^{36}$~erg/s) the most 
likely origin of the soft component is the emission from either a 
photoionized or collisionally heated diffuse gas in the binary system 
or from the NS surface. 
In the case of \xte,\ we found that BB emission 
from a relatively cold ($\sim$0.1~keV) and large ($\sim$100~km equivalent radius) 
region or from a much 
hotter ($\gtrsim$1~keV) and less extended ($<$100~m) spot, or, alternatively, emission 
from an optically thin gas (MKL) provided equally good fits to the data. 
While the discussion above and the rapid variability observed in the SFXT 
would argue against the presence of an accretion disk in these sources 
\citep[see also,][]{bozzo08}, we also consider that 
emission from a small and hot spot on the NS surface is inconsistent with  
models of accretion onto magnetic NS. 
The accretion flow in SFXT is, indeed, thought to 
be quasi-spherical, and expected to penetrate the NS magnetosphere mainly 
by means of the Rayleigh-Taylor and the Kelvin-Helmholtz instability 
\citep[see e.g.,][and references therein]{bozzo08}. 
In these circumstances, 
the size of the hot-spot over which the accretion takes place 
is expected to be inversely proportional to the X-ray luminosity and might cover  
a substantial fraction of the NS surface for 
$L_{\rm X}$$\lesssim$10$^{35}$~erg/s \citep[see][and references therein]{white83}.  

Taking these results into account, emission from an optically thin and 
diffuse gas around the NS seems to be a more reasonable explanation of the 
soft spectral component of \xte.\ 
According to this interpretation, the emitting region 
can be estimated from the normalization of the MKL component (see Table~\ref{tab:xtefit})
using the relation \citep[see e.g.,][and references therein]{zurita09b}:
$R_{\rm em}$=$\sqrt[3]{3 N_{\rm MKL}/10^{-14} (D/n_{\rm H})^2}$$\sim$10$^{13}$$a_{13}^{2/3}$~cm, where 
$N_{\rm MKL}$ is the normalization of the MKL component, D is the source distance,  
$n_{\rm H}$$\sim$$N_{\rm H}$/$a$, $a$ is the binary separation, and $a_{13}$=$a$/10$^{13}$~cm. 
Therefore, the radius $R_{\rm em}$ of the emitting region is compatible 
with the binary separation for a wide range of values of orbital periods 
similar to those measured in other SFXTs. We note also that the properties of this MKL component 
would be rather similar to those derived from the spectrum of the SFXT 
AX\,J1845.0-0433 \citep{zurita09b}. 

Another possibility that we investigated in Sect.~\ref{sec:xteresults} is  
the applicability of a model comprising a power law component and  
a partial covering to the X-ray spectrum of \xte.\ We concluded 
that this model can also provide a reasonably good fit to the data. 
A similar model was proposed to interpret the quiescent \XMM\ spectrum of the SFXT 
IGR\,J16207-5129 \citep{tomsick09}  
and might provide support for clumpy wind in these sources. In this case, one would 
expect part of the radiation from the NS to escape absorption by the clumps  
local to the source and be affected only by  interstellar absorption 
\citep[see, e.g.][]{walter07}.  
Finally, we showed that the COMPTT model also provided a reasonable fit to the \XMM\ spectra of 
\xte,\ and should thus be considered a valid alternative to the other models discussed above.  
We note, however, that neither the partial covering nor the COMPTT model could give an acceptable 
fit to the spectrum of \igrB\ (see also below). Given the similarities between the two sources, 
a spectral model that provides acceptable results in both cases should probably be favored 
(e.g, the CUTOFFPL+MKL model).

\subsection{\igrB\ }
\label{sec:igrbdiscussion}

The \XMM\ observation of \igrB,\ detected a very similar behavior to 
that discussed above for \xte.\ In the light curve of \igrB,\ a number of 
relatively small flares were observed to take place sporadically on a 
timescale of few thousands of seconds and were characterized by an 
X-ray flux $\sim$10-15 times higher than the underlying fainter 
persistent emission. The lowest X-ray flux that we measured from 
this source was $\sim$3$\times$10$^{-13}$~erg/cm$^2$/s  
(0.5-10~keV) and corresponds to a luminosity of 3.3$\times$10$^{32}$~erg/s 
(at a distance of 3~kpc), comparable to the value reported 
by \citet{kennea06}. The total dynamic range in the X-ray 
luminosity of \igrB\ between outburst and quiescence is thus 
$\gtrsim$10$^4$ (see also Sect.~\ref{sec:intro}). 

As for \xte,\ the hardness intensity diagrams and the rate 
resolved analysis carried out in Sect.~\ref{sec:results} showed that the  
variations in the X-ray flux measured during the \XMM\ observation  
were also accompanied by a significant change in the spectral properties 
of the source. In contrast to the case of \xte,\ the rate-resolved 
spectra of \igrB\ could not be fit by using a simple CUTOFFPL model. 
We showed that an acceptable fit to the data could, instead, be obtained by introducing an additional 
relatively cold ($kT$$\sim$0.08~keV) and extended ($\gtrsim$100~km) BB component, 
or a MKL model (see Tables~\ref{tab:igrbfit} and \ref{tab:igrbfittotal}).  
We note that the parameters measured for the MKL component are rather similar 
to those derived in the case of \xte.\ 
An equivalently good fit was provided by the BMC model. 
This model has the same number of free parameters as the CUTOFFPL+BB 
(see Table~\ref{tab:igrbfit}), and would predict similar properties for the 
temperature and the size in which the soft photons are produced 
\citep[the normalization of the BMC model is defined as the ratio of the source luminosity 
to the square of the distance in units of 10~kpc, see e.g.][and references therein]{sidoli09b}. 
Similar values of the fit parameters were also obtained from the analysis of the 
spectrum of \igrB\ extracted by using the total available exposure time of the \XMM\ 
observation (see Sect.~\ref{sec:igrresults} and Table~\ref{tab:igrbfittotal}). 

According to the discussion in Sect.~\ref{sec:xtediscussion}, a BB 
emission with these properties seems unlikely in the case of \igrB\ 
and thus we suggest that the CUTOFFPL+MKL model can provide a more reasonable description of 
the data. We note that in the \XMM\ spectrum of the supergiant HMXB IGR\,J16320-4751 
a similar soft component was found that could be fit with a BB of 0.07~keV but was 
attributed to a cloud surrounding the NS \citep{rodriguez06}. 
Following the CUTOFFPL+MKL interpretation, the rate resolved analysis carried out 
for the observation of \igrB\ would indicate that the properties of the MKL component 
do not change significantly with the source count rate and the increase in the 
hardness ratio observed in Figs.~\ref{fig:igrbhardness} and \ref{fig:igrb} is most likely due to a change 
in the CUTOFFPL photon index. Furthermore, no significant variations in the absorption column 
density were revealed in the different rate-resolved spectra. 
This is similar to the results discussed above for \xte.\ 

As for \xte,\ a comparison between the results of the present \XMM\ observation and the 
observations carried out in the same energy band (0.5-10~keV) 
when this source was in outburst does not indicate a clear correlation 
between the power law photon index, the absorption column density, and the source 
X-ray flux (see Sect.~\ref{sec:intro}). It is interesting that 
the soft component in this source detected by the \XMM\ observation appears 
to have a different origin from that detected by \citet{sidoli09b} when \igrB\ was in outburst 
(see Sect.~\ref{sec:intro}). 
On that occasion, the soft component appeared to be caused by thermal  
emission from a hot-spot on the NS surface. We note that, even if the soft 
component observed by \XMM\ is interpreted in terms of a BB emission, 
the emitting region derived from the fit is considerably larger 
than the NS radius 
and it is thus unlikely that it is produced on the NS surface. \\

Finally, for both \xte\ and \igrB,\ we investigated whether  
the harder spectral component detected in these sources might be produced by the 
X-ray emission from the NS supergiant companion.  
The time-averaged X-ray luminosity that we measured from these sources in quiescence 
matches quite well the luminosity expected from an isolated OB supergiant or from colliding winds 
in a binary containing OB supergiant stars \citep[see e.g.,][for a review]{gudel09}. 
However, this interpretation appears to be contrived
for the following reasons.  
The X-ray spectrum of isolated or colliding wind binaries   
with OB supergiant stars is usually described well by a model 
comprising one or more thermal components \citep[MKL in {\sc xspec}, see e.g.,][]{gudel09}.
The softer MKL component has a typical temperature of $\sim$0.2-0.7~keV, and is thus 
similar to those we detected in \igrB\ and \xte.\ 
This component is thought to be generated by  
the shocks within the stellar wind. The hotter MKL component, extending up to 
several keV, can have a temperature as high as $\sim$1-3~keV and is characterized by  
a number of very prominent emission lines \citep[see also][]{raassen08}.
This hard component is usually interpreted in terms of magnetically confined 
wind shocks, highly compressed wind shocks, or inverse Compton scattering 
of photospheric UV photons by relativistic 
particles accelerated within the shocks \citep{colombo07}. 
Possible detections of a non-thermal X-ray emission 
from OB supergiant stars were reported only in only two cases, 
but they still lack confirmation \citep{gudel09}. 

The X-ray spectra of \xte\ and \igrB\ were reproduced well using a 
CUTOFFPL model, and no prominent emission lines were detected. 
The values of the photon index, $\Gamma$, derived 
from \XMM\ are also comparable to those  
obtained previously when the sources were in outburst, thus  
suggesting that a common mechanism produces their harder X-ray component. 
Furthermore, the relatively rapid X-ray variability (of period few thousands seconds)  
observed in the lightcurves of \igrB\ and \xte\ is not reminiscent of the typical 
X-ray variability of the OB stars, which, when present, 
takes place on longer timescales  
\citep[tens of ks, see e.g,][]{colombo07}. 

We conclude that the harder X-ray emission from \xte\ and \igrB\ is most likely 
due to residual accretion taking place onto the NS at a much lower  
rate than during outburst.

\section{Conclusions}
\label{sec:conclusion}

The three \XMM\ observations that we have analyzed in the present work, indicate that the quiescent 
spectra of the two  prototypical SFXT \xte\ and \igrB\ are characterized by 
two different spectral components, one dominating the spectrum at the softer energies 
($\lesssim$2~keV) and the other one being more prominent above 2~keV. 

The properties of the soft component ($\lesssim$2~keV) could be reasonably well constrained 
in the case \igrB,\ where the absorption 
column density was relatively low ($<$10$^{22}$~cm$^{-2}$), whereas 
in the case of \xte\ the detection of this component is less significant.  
However, the similarity in the timing and spectral behavior observed in the quiescent 
state of the two sources argues in favor of adopting the same spectral model 
for both of them. 
We suggested that the model comprising a CUTOFFPL 
component at the higher energies plus a MKL component would provide 
a reasonable description of the data and a plausible physical explanation 
of the properties observed in the two sources (Sect.~\ref{sec:discussion}).  
According to this interpretation, the MKL component would represent the contribution 
to the total X-ray emission of the shocks 
in the wind of the supergiant companion.
The results of the fits with this model to the data of 
the three observations inferred a temperature of the MKL component 
and an emitting region 
comparable with the values found also in the case 
of the SFXT AX\,J1845.0-0433. 

Similar soft spectral components have been detected in many other HMXBs and SGXBs.
In a few cases, the detection of a number of prominent emission lines in 
the high resolution X-ray spectra of these sources carried out with the 
gratings onboard \chan\ and the RGS onboard \XMM\ 
\citep[see e.g.,][]{watanabe06} have convincingly demonstrated 
that these components are produced by the stellar  
wind around the NS, and proved to be a powerful diagnostic 
to probe the structure and composition of the stellar 
wind in these systems. 
The statistics of the present \XMM\ observations 
is far too low to permit a similar in-depth study of the stellar wind in 
the case of \xte\ and \igrB.\ 
Furthermore, because of the relatively low luminosity and the high absorption 
that characterize the emission of these sources in quiescence, 
observations at higher spectroscopic resolution are probably 
too challenging for the present generation of X-ray satellites, and  
the improved sensitivity of the X-ray spectrometers 
planned for future X-ray missions (e.g. IXO) 
is probably required to firmly establish the presence of a soft 
spectral component in the quiescent emission of the SFXT sources 
and shed light on its nature. 

If the harder X-ray emission (2-10~keV) detected from the \XMM\ observations 
of \xte\ and \igrB\ was really produced by residual accretion as we argued 
in the previous section, then the accretion process in these sources  
would take place over more than 4 orders of magnitude of X-ray luminosity\footnote{That 
we did not find any evidence for X-ray eclipses in 
the three \XMM\ observations that we analyzed
(this is unlike the case of IGR\,J16479-4514, see Sect.~\ref{sec:intro}) is also consistent 
with this interpretation.}. 
This is similar to the results reported for the SFXT IGR\,J17544-2619 
\citep{rampy09} and, possibly, for the SFXT SAX\,J1818.6-1703 
\citep[in the latter case the origin of the lowest quiescent emission 
remains unclear,][]{bozzo09}. 
In the case of IGR\,J17544-2619, \citet{rampy09} ascribed the high dynamic range in the 
X-ray luminosity to the accretion of clumps from the wind of the supergiant star 
with a high density contrast with respect to the surrounding homogeneous wind. 
However, it was also suggested that a similar variability might 
result from the transition across different accretion 
regimes onto the NS \citep{bozzo08}.  
We note that a similar scenario can be envisaged for interpreting the variations in the X-ray flux 
observed during the multiple small flares detected in the present observations. 
Even though they took place at a much lower  
luminosity level than the brightest outbursts (a factor of $\sim$10$^3$-10$^4$), our analysis 
showed that all these events shared a number of similar timing and spectral properties. In particular, the timescales 
on which the smaller flares develop is comparable with the decay timescale of the source luminosity 
during the outbursts (see Sect.~\ref{sec:intro}), and the spectral photon indices and absorption 
column densities measured from the \XMM\ observations are also in qualitative agreement with those reported 
previously when the sources were observed at a much higher X-ray luminosity level 
(see Sect.~\ref{sec:intro}). It is thus most likely that the transitions 
between the lower quiescent states and the small flares detected by \XMM\ from \xte\ and \igrB\ 
might have been triggered by the same mechanism that sometimes gives rise to the brightest outbursts 
(i.e., the accretion of clumps from the stellar wind and/or the transition between different 
accretion regimes of the NS, see Sect.~\ref{sec:intro}). 

In contrast to the case for the SFXT IGR\,J18483-0311, we did not detect any pulsation in the 
quiescent emission of either \xte\ or \igrB,\ and provided in Sects.~\ref{sec:xteresults} and \ref{sec:igrresults} 
the corresponding upper limits to the spin periods and pulsed fractions we were able to infer from  
the present data. 

Deep pointed observations of SFXTs in quiescence are still 
required in order to understand the origin of the peculiar X-ray variability 
of these sources and distinguish between different models proposed to 
interpret their behavior.

\section*{Acknowledgments}
EB thanks N. Schartel and the \XMM\ staff for the rapid schedule of the 
\XMM\ observation of \xte\ after the outburst occurred on 
2009 March 10, and R. Farinelli for helpful discussions. 
We thank the anonymous referee for useful comments.

{}

\end{document}